\documentclass[aps,prb,longbibliography,reprint,superscriptaddress,citeautoscript]{revtex4-1}

\usepackage[applemac]{inputenc}
\usepackage[T1]{fontenc}
\usepackage[american]{babel}
\usepackage[scaled]{helvet}
\usepackage{graphicx,amssymb,bm,mathtools,textcomp,courier,multirow,bigdelim,color,hyperref}
\usepackage[normalem]{ulem}	

\newcommand{\Dirac}[3]{\left\langle #1 \left| #2\right| #3\right\rangle}

\newcommand{\op}[2]{\left.{\left| #1\right\rangle\left\langle #2\right|}\right.}

\newcommand{\ket}[1]{\left|\left. #1 \right\rangle\right.}
\newcommand{\bra}[1]{\left\langle\left. #1 \right|\right.}

\newcommand{\abs}[1]{\left| #1 \right|}

\newcommand{\figureshortname}{Fig.}
\newcommand{\equationshortname}{Eq.}
\newcommand{\tableshortname}{Tab.}
\newcommand{\tu}[1]{\textsuperscript{#1}}

\newcommand{\eref}[1]{\equationshortname~\eqref{#1}}
\newcommand{\sref}[1]{Sec.~\ref{#1}}
\newcommand{\aref}[1]{Appx.~\ref{#1}}
\newcommand{\cref}[1]{Chapter~\ref{#1}}
\newcommand{\fref}[1]{\figureshortname~\ref{#1}}
\newcommand{\tref}[1]{\tableshortname~\ref{#1}}
\newcommand{\rcite}[1]{Ref.~[\onlinecite{#1}]}
\newcommand{\mrcite}[2]{Refs.~[\onlinecite{#1},\onlinecite{#2}]}
\newcommand{\mmrcite}[3]{Refs.~[\onlinecite{#1},\onlinecite{#2},\onlinecite{#3}]}

\graphicspath{{Pictures/}}

\begin{document}
\def\sectionautorefname{Sec.}

\title{Quantum computation with three-electron double quantum dots at an optimal operation point}

\author{Sebastian Mehl}
\email{s.mehl@fz-juelich.de}
\affiliation{JARA-Institute for Quantum Information, RWTH Aachen University, D-52056 Aachen, Germany}
\affiliation{Peter Gr\"unberg Institute (PGI-2), Forschungszentrum J\"ulich, D-52425 J\"ulich, Germany}

\date{\today}

\begin{abstract}
The author analyzes quantum computation with the hybrid qubit (HQ) that is encoded using the three-electron configuration of a double quantum dot. All gate operations are controlled with electric signals, while the qubit remains at an optimal operation point that is insensitive to noise. An effective single-qubit description is derived, and two-qubit interactions are suggested using Coulomb and exchange interactions. Universal quantum control is described and numerically simulated using realistic parameters for HQs in Si and GaAs. High-fidelity quantum computing at the threshold of quantum error correction is possible if the Coulomb interactions between the HQs stay weak.
\end{abstract}

\maketitle

\section{Introduction}

Spin qubit quantum computers are promising platforms to achieve quantum computation.\cite{loss1998,kane1998} The electron spin of a gate-defined quantum dot (QD) naturally defines a two-level quantum system that encodes one bit of quantum information \cite{loss1998,awschalom2002}. Also a duo\cite{levy2002,taylor2005}, a trio\cite{divincenzo2000}, and a quartet\cite{bacon2000,kempe2001} of such singly occupied QDs have been proposed as realizations of quantum bits. The so-called hybrid qubit (HQ) is encoded using a double QD (DQD) that is occupied with three electrons\cite{shi2012}. The name HQ was introduced because the HQ is either a spin or a charge qubit depending on its operation principle. A spin qubit is well protected from charge noise, but typical spin qubits cannot be operated all-electrically. A charge qubit can be manipulated with electric fields, but it is therefore also susceptible to charge noise.

The HQ can be operated exclusively with electric signals similar to the triple QD (TQD) qubit that is coded using three singly occupied QDs.\cite{divincenzo2000,shi2012} Because subnanosecond controls of the electrostatic potentials of QDs can be realized \cite{petta2005,maune2012}, electric manipulations of QD qubits are always more favorable than magnetic manipulations. Even though single spins can also be controlled with magnetic field pulses,\cite{koppens2006,pla2012} it is difficult to selectively control a single spin with magnetic fields when this spin is in the vicinity of other spins.\cite{veldhorst2014} Electric fields can also modify single spins indirectly (e.g. via spin-orbit interactions or magnetic field gradients), but high-fidelity gates still remain challenging.\cite{nowack2007,kawakami2014,yoneda2014}

Most conveniently, the HQ is manipulated by the transfer between the $\left(n_{\text{QD}_{1}},n_{\text{QD}_{2}}\right)=\left(1,2\right)$ and $\left(2,1\right)$ configurations,\cite{shi2012,shi2013,shi2014,kim2014,kim2014-2} similar to a charge qubit.\cite{petersson2010,dovzhenko2011,cao2013} $n_{\text{QD}_{1}}$ and $n_{\text{QD}_{2}}$ are the electron numbers at QD$_1$ and QD$_2$. These manipulations require precise, subnanosecond pulses of electric signals with large amplitudes. Considerable advances for the HQ were possible using an asymmetric DQD configuration.\cite{shi2014,kim2014,kim2014-2} For a doubly occupied QD, the addition of the second electron usually requires higher energy in the triple configuration compared to the singlet configuration. DQDs were constructed, where this singlet-triplet energy difference is small for one of the QDs, but it is much larger for the other one.

This paper studies DQDs with asymmetric addition energies in the layout of \fref{fig:01}. QD$_2$ always has a small singlet-triplet energy difference, but the singlet-triplet energy difference difference is large for QD$_1$. It will be shown that the HQ has a small energy difference in the $\left(1,2\right)$ configuration. When approaching the $\left(2,1\right)$ configuration, the qubit states pass through two avoided level crossings. It has been described theoretically\cite{shi2012} and shown experimentally \cite{shi2014,kim2014} that these two anticrossings are sufficient for the single-qubit control of the HQ with the appropriate tuning protocols through the anticrossings.

It is also possible to operate the HQ in the vicinity of an anticrossing, where all the qubit operations are realized with microwave pulses of small amplitudes.\cite{kim2014-2} An anticrossing is a sweet spot in the energy diagram where low-frequency noise in the control parameter does not dephase the qubit. Such sweet-spot operations have improved the coherence times for superconducting qubits \cite{vion2002,koch2007,schreier2008}. Note that microwave gates have also been realized for other spin qubit encodings, as for DQDs\cite{shulman2014} and TQDs\cite{taylor2013,medford2013}.

For HQs, the entangling operations remain more challenging than the single-qubit gates. \rcite{shi2012} proposed that strong electrostatic couplings can be used to entangle two HQs. Also the rapid transfer of electrons between the HQs has been studied.\cite{mehl2015-1} In this paper, I show that the HQ can always be operated at its sweet spot while weak couplings between HQs enable two-qubit gates. I quantify the inter-qubit couplings with Coulomb and exchange interactions, and show that both interactions enable two-qubit gates.

The main findings of this paper are explicit manipulation protocols for HQs of high fidelities, while all the HQs are operated at their sweet spots. I simulate the gate operations with typical qubit parameters and show that these gates tolerate the dominant noise sources of HQs. At the sweet spot, the HQ keeps enough character of a spin qubit to be protected from charge noise, but it is already sufficiently close to a charge qubit to be rapidly tunable. To employ these gates experimentally, one has to build setups that have weak Coulomb interactions between the HQs at their sweet spots.

The organization of the paper is as follows. \sref{sec:Single} introduces the description of the HQ, and it specifies the magnitudes of the relevant parameters. The single-qubit control, the readout, and the initialization are also described. \sref{sec:TwoQubit} derives the effective Hamiltonians for the couplings between HQs that are generated from Coulomb and exchange interactions. \sref{sec:UnivCont} specifies two approaches to realize universal control of the HQ. Not only entangling operations are discussed, but it is also shown that single-qubit gates are possible. \sref{sec:Noise} discusses the influence of electric and magnetic noise for HQs that are operated at their sweet spots. \sref{sec:Summ} summarizes the findings.

\section{\label{sec:Single}
Single-Qubit Description}

\begin{figure}
\includegraphics[width=0.3\textwidth]{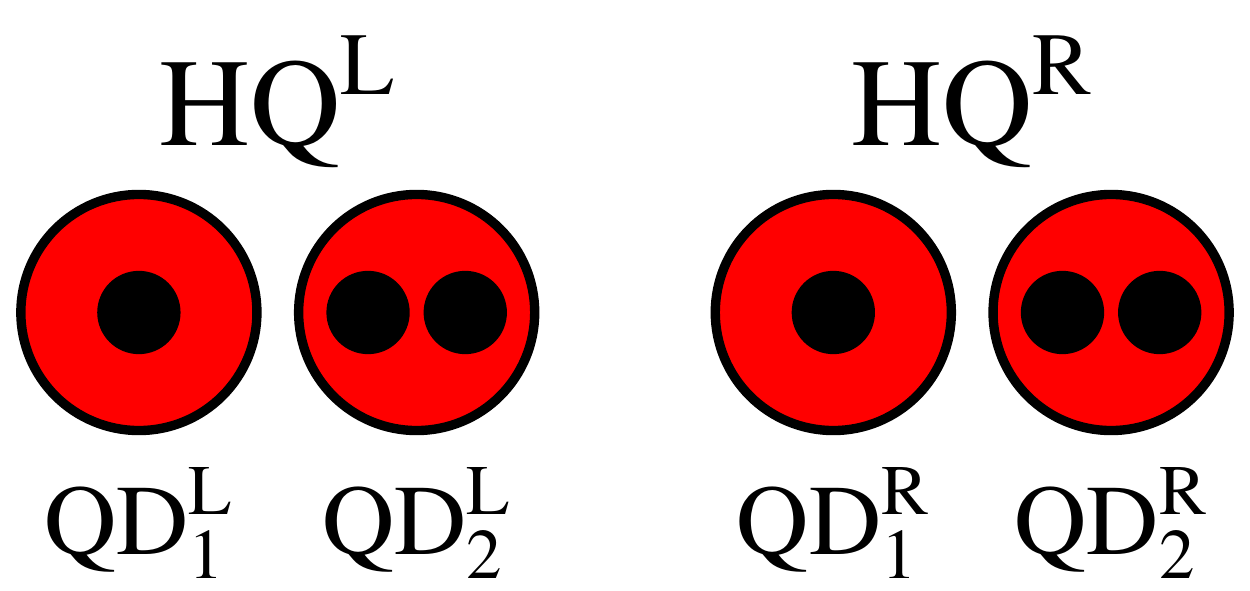}
\caption{\label{fig:01}
Setup of two HQs. Each HQ is coded using the three electron configuration of a DQD. The superscripts L and R label the positions of the HQs; the subscripts 1 and 2 label the QDs.
}
\end{figure}

The HQ is coded using a DQD that is occupied with three electrons, as shown in \fref{fig:01}. The qubit manipulations require a much larger singlet-triplet energy splitting of a doubly occupied QD for QD$_1$ compared to QD$_2$. I consider only the subspace encoding in the $S=1/2$, $s_z=1/2$ spin configuration\cite{divincenzo2000,mehl2013-1}. Even though a global magnetic field is not required to realize single-qubit gates, it will be necessary to apply magnetic fields to initialize a HQ to $S=1/2$, $s_z=1/2$.\footnote{Note that the tunnel couplings between HQs couple between spin subspaces such that two-qubit gates can be problematic if the two HQs are initialized to different spin subspaces.}
QD$_1$ can be used for the initialization of a singlet because a two-electron configuration quickly equilibrates to a singlet under thermal relaxation. In the presence of an external magnetic field, also the spin-up configuration at QD$_2$ can be initialized because it has lower energy than the spin-down configuration.\footnote{
I define that \protect{$\uparrow$} is the spin configuration with the lowest energy. Note that there are materials of positive g factors (e.g. Si) and negative g factors (e.g. GaAs).}

The HQ is described in the basis of the lowest states with $S=1/2$, $s_z=1/2$. One considers $\ket{\text{x}}=\ket{\text{S}\uparrow}$ in the (2,1) configuration, and the states $\ket{1}\propto\sqrt{2}\ket{\downarrow \text{T}_+}-\ket{\uparrow \text{T}_0}$ and $\ket{0}=\ket{\uparrow \text{S}}$ in the (1,2) configuration. The singlet $\ket{\text{S}}\propto\ket{\uparrow\downarrow}-\ket{\downarrow\uparrow}$ and the triplets $\ket{\text{T}_+}=\ket{\uparrow\uparrow}$, $\ket{\text{T}_0}\propto\ket{\uparrow\downarrow}+\ket{\downarrow\uparrow}$, and $\ket{\text{T}_-}=\ket{\downarrow\downarrow}$ describe the two-electron configurations at a QD. The three-electron states with $S=1/2$, $s_z=1/2$ are defined using the standard spin addition rules \cite{sakurai1994}. The Hamiltonian for the HQ in the basis $\left\{\ket{\text{x}},\ket{1},\ket{0}\right\}$ is
\begin{align}
\mathcal{H}_{\left\{\ket{\text{x}},\ket{1},\ket{0}\right\}}=\left(\begin{array}{ccc}
\epsilon & -\Delta_1 & \Delta_0\\
-\Delta_1 & \delta & 0\\
\Delta_0 & 0 & 0
\end{array}
\label{eq:Ham1}
\right).
\end{align}

The Coulomb repulsion is high for a doubly occupied QD in the triplet configuration, which raises the energy of $\ket{1}$ compared to $\ket{0}$ by $\delta$. $\ket{\text{x}}$ and $\ket{1}$ ($\ket{\text{x}}$ and $\ket{0}$) are coupled via the real parameter $\Delta_1>0$ ($\Delta_0>0$), which is called the tunnel coupling in the following. $\epsilon$ is the detuning parameter between the $\left(2,1\right)$ and $\left(1,2\right)$ configurations. The electron configuration in $\left(2,1\right)$ [$\left(1,2\right)$] has lower energy for $\epsilon<0$ [$\epsilon>0$], and $\left(2,1\right)$ and $\left(1,2\right)$ have equal energies at $\epsilon=0$.

\begin{figure}
\includegraphics[width=0.49\textwidth]{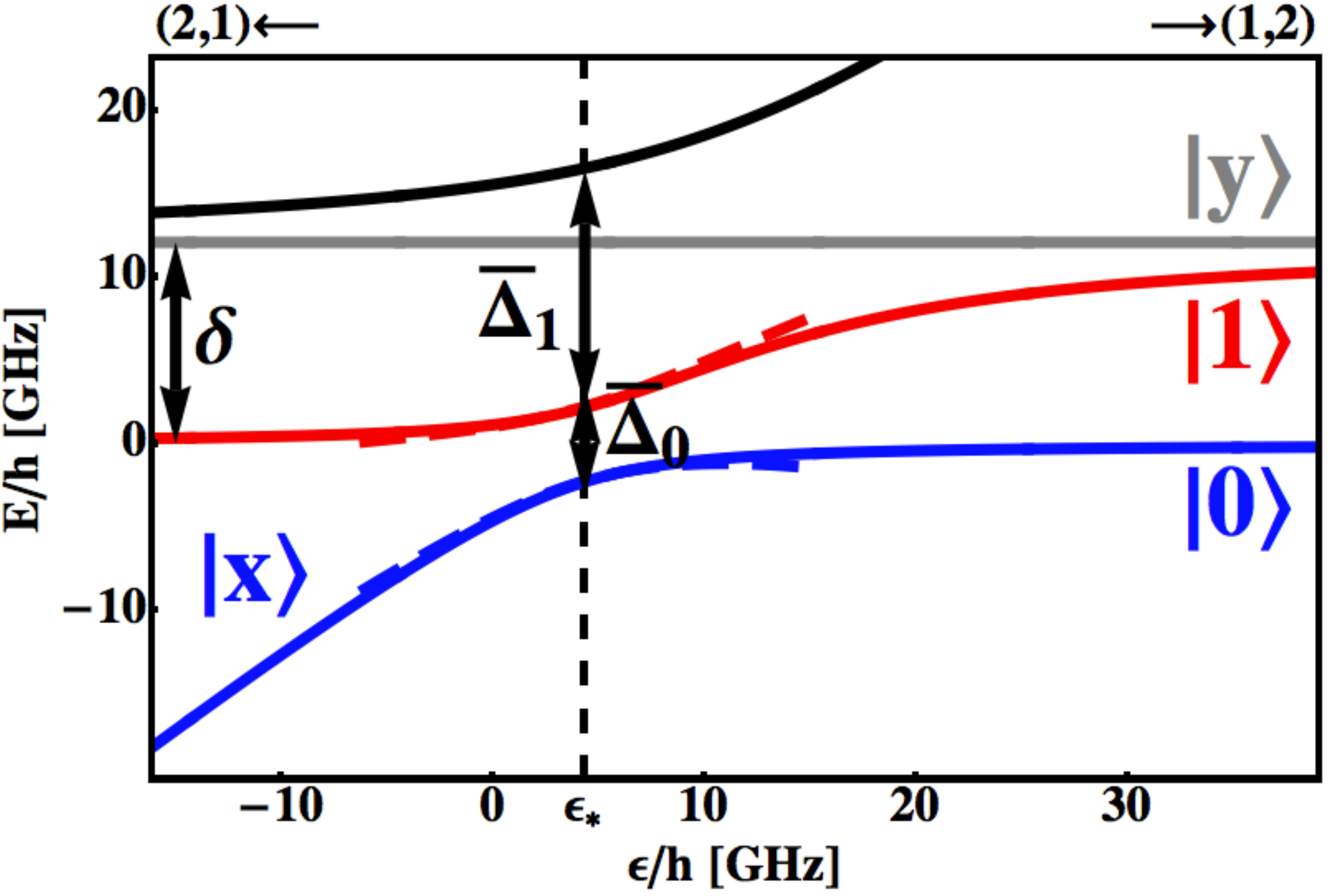}
\caption{\label{fig:02}
Energy diagram of the relevant states to describe the HQ in the $S=1/2$, $s_z=1/2$ spin configuration of a three-electron DQD, according to \eref{eq:Ham1}, for $2\Delta_1/h=14.5~\text{GHz}$, $2\Delta_0/h=5.2~\text{GHz}$, and $\delta/h=12.1~\text{GHz}$. $\epsilon$ models the energy detuning between the $\left(2,1\right)$ and $\left(1,2\right)$ charge configurations. The HQ is coded using the two states with the lowest energies. The HQ is in $\left(1,2\right)$ for $\epsilon\rightarrow\infty$ with the excited state $\ket{1}\propto\sqrt{2}\ket{\downarrow \text{T}_+}-\ket{\uparrow \text{T}_0}$ and the ground state $\ket{0}=\ket{\uparrow \text{S}}$. Only the ground state $\ket{\text{x}}=\ket{\text{S}\uparrow}$ is in $\left(2,1\right)$ for $\epsilon\rightarrow -\infty$, while the excited state remains in $\left(1,2\right)$. There are two state couplings ($\ket{1}\leftrightarrow\ket{\text{x}}$ and $\ket{0}\leftrightarrow\ket{\text{x}}$) at $\epsilon\sim 0$. The gray line shows the $S=3/2$, $s_z=1/2$ state (called $\ket{\text{y}}$) in $\left(1,2\right)$ that is uncoupled to $\left(2,1\right)$. All gate operations for the HQ are realized at the sweet spot $\epsilon_*$, where the qubit is insensitive to small variations in $\epsilon$. The dashed lines approximate the qubit levels in the vicinity of $\epsilon_*$ [see \eref{eq:Ham3}]. The parameters $\overline{\Delta}_1$ and $\overline{\Delta}_0$ are defined in the text.
}
\end{figure}

\eref{eq:Ham1} describes a three level system with two anticrossings. \fref{fig:02} shows the energy levels for typical qubit parameters in the transition region between $\left(2,1\right)$ and $\left(1,2\right)$. The two states with the lowest energies encode the HQ. For $\epsilon\ll0$, only the ground state is in $\left(2,1\right)$, while the first excited state is in $\left(1,2\right)$. This configuration is ideal for the readout of the HQ because a nearby charge detector can distinguish the two states of the HQ, similar to the readout of a singlet-triplet qubit \cite{reilly2007,barthel2009,barthel2010-2}. For $\epsilon\gg0$, the HQ is in $\left(1,2\right)$ and the energy difference between the qubit states is unchanged when $\epsilon$ is slightly modified. This regime is highly protected from charge noise \cite{hu2006,dial2013}. Note that there is additionally one $S=3/2$, $s_z=1/2$ state in the $\left(1,2\right)$ configuration (called $\ket{y}$) that is uncoupled to the $\left(2,1\right)$ configuration.

I consider a HQ with similar magnitudes of the singlet-triplet energy splitting at QD$_2$ $\delta$ and the tunnel couplings $\Delta_1$ and $\Delta_0$: $\delta\gtrsim\Delta_1, \Delta_0$. The singlet-triplet energy splitting at QD$_1$ is much larger than $\delta$ such that doubly occupied triplets at QD$_1$ are neglected. Note that such HQs have been realized using Si DQDs,\cite{shi2014,kim2014,kim2014-2} where the singlet-triplet energy difference depends on the orbital and the valley energy splittings. The orbital energy splitting is determined by the confining potential of a QD, while the valley energy splitting is determined by the potential landscape at the atomic scale. A recent experiment showed that the valley splitting of a Si QD can be controlled electrically \cite{yang2013}. Note that HQs can also be build using GaAs QDs, but these setups necessarily require asymmetric DQD potentials to operate them with the equivalent tuning pulses. In this case, the smaller QD has a large singlet-triplet splitting; the larger or elliptic QD has a small singlet-triplet splitting.\cite{mehl2013-2,mehl2014-2,hiltunen2015} The parameters of previous experiments with Si HQs are used in the following: $2\Delta_1/h=14.5~\text{GHz}$, $2\Delta_0/h=5.2~\text{GHz}$, and $\delta/h=12.1~\text{GHz}$ (cf. \mmrcite{shi2014}{kim2014}{kim2014-2}).

$\Delta_1$ in \eref{eq:Ham1} is caused by the coupling to an exited orbital, which is larger than the ground state orbital that determines $\Delta_0$. Tunnel couplings are exponentially suppressed with the distance between localized orbitals. Therefore, $\Delta_1$ is significantly larger than $\Delta_0$, and it is meaningful to analyze \eref{eq:Ham1} in the eigenbasis of the larger anticrossing. The states are rotated to 
$\ket{\widetilde{\text{x}}}=
\cos\left(\theta_\epsilon/2\right)\ket{\text{x}}+
\sin\left(\theta_\epsilon/2\right)\ket{1}$ and 
$\ket{\widetilde{1}}=
-\sin\left(\theta_\epsilon/2\right)\ket{\text{x}}+
\cos\left(\theta_\epsilon/2\right)\ket{1}$, with 
$\sin\left(\theta_\epsilon\right)=-2\Delta_1/U_\epsilon$,
$\cos\left(\theta_\epsilon\right)=(\epsilon-\Delta)/U_\epsilon$, and
$U_\epsilon=\sqrt{\left(2\Delta_1\right)^2+\left(\epsilon-\delta\right)^2}$.
This basis rotation modifies \eref{eq:Ham1} to
\begin{align}
\label{eq:Ham2}
\mathcal{H}&_{\left\{\ket{\widetilde{\text{x}}},\ket{\widetilde{1}},\ket{0}\right\}}=
\\\nonumber&
\left(\begin{array}{ccc}
\frac{\epsilon+\delta}{2}+\frac{U_\epsilon}{2} & 0 & \Delta_0\cos\left(\frac{\theta_\epsilon}{2}\right)\\
0 & \frac{\epsilon+\delta}{2}-\frac{U_\epsilon}{2} & -\Delta_0\sin\left(\frac{\theta_\epsilon}{2}\right)\\
\Delta_0\cos\left(\frac{\theta_\epsilon}{2}\right) & -\Delta_0\sin\left(\frac{\theta_\epsilon}{2}\right) & 0
\end{array}
\right).
\end{align}

The coupling of $\ket{\widetilde{\text{x}}}$ to $\left\{\ket{\widetilde{1}},\ket{0}\right\}$ is neglected in \eref{eq:Ham2} to describe the low-energy subspace that defines the HQ. It will be shown that the latter subspace has an anticrossing at $\epsilon_*= \Delta_1^2/\delta$, while $U_{\epsilon_*}\gg\Delta_0$. The two-level system $\left\{\ket{\widetilde{1}},\ket{0}\right\}$ is approximated by:
\begin{align}
\mathcal{H}_{\left\{\ket{\overline{1}},\ket{\overline{0}}\right\}}&\approx
\frac{\overline{\Delta}_0}{2}\sigma_z+
\mathcal{A}\left(\epsilon\right)\sigma_x,
\label{eq:Ham3}
\end{align}
with $\overline{\Delta}_0=2\Delta_0\sin\left(\theta_*/2\right)$,
$\mathcal{A}\left(\epsilon\right)=
\frac{\delta^2}{2\left(\delta^2+\Delta_1^2\right)}\left(\epsilon-\epsilon_*\right)$, and $\theta_*=\theta_{\epsilon_*}$.
To derive \eref{eq:Ham3}, the dependency of $\sin\left(\theta_\epsilon/2\right)$ on $\epsilon$ is neglected because this term varies slowly in the vicinity of $\epsilon_*$. Then the qubit's energy splitting $
\sqrt{
\left(\frac{\epsilon+\delta-U_\epsilon}{4}\right)^2+
\left[\Delta_0\sin\left(\theta_*\right)\right]^2
}$ within $\left\{\ket{\widetilde{1}},\ket{0}\right\}$ is minimized with respect to $\epsilon$, giving the first term in \eref{eq:Ham3}, 
with $\overline{\Delta}_0=2\Delta_0\sin\left(\theta_*/2\right)$ for 
$\epsilon_*\approx \Delta_1^2/\delta$. The second term in \eref{eq:Ham3} is obtained from $
\left.\sqrt{
\partial_\epsilon\left(\frac{\epsilon+\delta-U_\epsilon}{4}\right)^2+
\partial_\epsilon\left[\Delta_0\sin\left(\theta_*\right)\right]^2
}\right|_{\epsilon=\epsilon_*}$, giving $\frac{\delta^2}{2\left(\delta^2+\Delta_1^2\right)}$.

The basis in \eref{eq:Ham3} has been rotated to
\begin{align}
\ket{\overline{1}}&=\frac{1}{\sqrt{2}}\left[
-\sin\left(\frac{\theta_*}{2}\right)\ket{\text{x}}
+\cos\left(\frac{\theta_*}{2}\right)\ket{1}
+\ket{0}\right],\\
\ket{\overline{0}}&=\frac{1}{\sqrt{2}}\left[
-\sin\left(\frac{\theta_*}{2}\right)\ket{\text{x}}
+\cos\left(\frac{\theta_*}{2}\right)\ket{1}
-\ket{0}\right],
\end{align}
with the definitions of the Pauli operators 
$\sigma_z=\op{\overline{1}}{\overline{1}}-\op{\overline{0}}{\overline{0}}$ 
and 
$\sigma_x=\op{\overline{1}}{\overline{0}}+\op{\overline{0}}{\overline{1}}$.
Note that $\ket{\overline{1}}$ and $\ket{\overline{0}}$ have finite contributions in $\left(2,1\right)$ and $\left(1,2\right)$.
$\theta_*\approx 1.34\pi$ is the mixing angle, with $\cos^2\left(\theta_*/2\right)\approx0.26$ and $\sin^2\left(\theta_*/2\right)\approx0.74$. \fref{fig:02} proofs that the effective two level system in \eref{eq:Ham3} describes the HQ close to $\epsilon_*$. The leakage state $\ket{\widetilde{\text{x}}}$ is raised by
$\overline{\Delta}_1=\delta+\Delta_1^2/\delta$.
$\overline{\Delta}_0=2\Delta_0\sin\left(\theta_*/2\right)$ is the energy difference between $\ket{\overline{1}}$ and $\ket{\overline{0}}$.

\eref{eq:Ham3} permits the usual Rabi control of the qubit with microwave drives of the detuning parameter $\epsilon$ \cite{slichter1990,vandersypen2005}. If $\epsilon$ is driven with small amplitudes around $\epsilon_*$
[$\mathcal{A}\left(\epsilon\right)
\rightarrow
\mathcal{A}\cos\left(2\pi\Omega t/h+\phi\right)$], then all possible single-qubit operations can be realized for $\Omega=\overline{\Delta}_0$ when the phase $\phi$ is varied. \eref{eq:Ham3} gives in the rotating frame with $
\frac{\overline{\Delta}_0}{2}\sigma_z$ the static Hamiltonian
\begin{align}
\mathcal{H}^{\text{rwa}}=
\mathcal{A}\left[
\cos\left(\phi\right)\sigma_x+\sin\left(\phi\right)\sigma_y\right].
\label{eq:Ham4}
\end{align}
\eref{eq:Ham4} uses implicitly a rotating wave approximation, which is valid for $\overline{\Delta}_0\gg\mathcal{A}$.

\section{\label{sec:TwoQubit}
Two-Qubit Interactions}

This section describes the interactions between two HQs. The superscripts L and R label the positions of the HQs. QD$_2^{\text{L}}$ and QD$_1^{\text{R}}$ are the neighboring QDs from HQ\tu{L} and HQ\tu{R} (cf. \fref{fig:01}).

\subsection{\label{sec:Cap}
Capacitive Coupling}

The charge configurations of two HQs couple capacitively via the Coulomb interaction. The dominant contribution is determined by the electron configurations at QD$_2^{\text{L}}$ ($n_{\text{QD}_2^{\text{L}}}$) and  QD$_1^{\text{R}}$ ($n_{\text{QD}_1^{\text{R}}}$) according to
\begin{align}
\mathcal{C}=\kappa~n_{\text{QD}_2^{\text{L}}}n_{\text{QD}_1^{\text{R}}}.
\label{eq:Coul1}
\end{align}
The magnitude of the coupling parameter $\kappa$ depends on the layout of the experiment and the QD material. $\kappa$ can be large; e.g., one can approximate its magnitude by the Coulomb interaction of two electric point charges that are $250~\text{nm}$ apart, giving $\kappa\approx\frac{e^2}{4\pi\epsilon_0\epsilon_r}\frac{1}{250~\text{nm}}\approx 500~\mu\text{eV}$
for the dielectric constant $\epsilon_r=11.7$ of silicon \cite{ioffe}. This value agrees with the approximation of a few tenth of meV in \rcite{koh2012}.

The entangling operations that follow require much smaller $\kappa$ than the naive estimate given above. The environment around the QDs\cite{trifunovic2012} and the metallic gates partly screen the capacitive couplings between the HQs, and thus they reduce $\kappa$. Also the layout of the four QDs can be be designed such that the Coulomb interactions between neighboring HQs are lowered.\cite{srinivasa2014} Finally, bringing QD$_2^{\text{L}}$ and  QD$_1^{\text{R}}$ further apart lowers $\kappa$.

For two HQs operated at their sweet spots, $\epsilon_*^{\text{L}}$ and $\epsilon_*^{\text{R}}$, the projection of \eref{eq:Coul1} to the subspace $\left\{
\ket{\overline{1}^{\text{L}}\overline{1}^{\text{R}}},
\ket{\overline{1}^{\text{L}}\overline{0}^{\text{R}}},
\ket{\overline{0}^{\text{L}}\overline{1}^{\text{R}}},
\ket{\overline{0}^{\text{L}}\overline{0}^{\text{R}}}\right\}$ gives the two-qubit interaction
\begin{align}
\mathcal{C}_{2Q}=
\mathcal{X}_{\mathcal{C}}
\sigma_x^{\text{L}}\sigma_x^{\text{R}},
\label{eq:Coul2}
\end{align}
with $\mathcal{X}_{\mathcal{C}}=-\frac{\kappa}{4}
\sin^2\left(\theta^{\text{L}}_*/2\right)
\sin^2\left(\theta^{\text{R}}_*/2\right)$. \eref{eq:Coul2} lowers the energies of the configurations 
$\ket{\overline{1}^{\text{L}}\overline{1}^{\text{R}}}+\ket{\overline{0}^{\text{L}}\overline{0}^{\text{R}}}$ and 
$\ket{\overline{0}^{\text{L}}\overline{1}^{\text{R}}}+\ket{\overline{1}^{\text{L}}\overline{0}^{\text{R}}}$ compared to 
$\ket{\overline{1}^{\text{L}}\overline{1}^{\text{R}}}-\ket{\overline{0}^{\text{L}}\overline{0}^{\text{R}}}$ and
$\ket{\overline{0}^{\text{L}}\overline{1}^{\text{R}}}-\ket{\overline{1}^{\text{L}}\overline{0}^{\text{R}}}$. The first states have large weights in $\left(2,1,2,1\right)$ or $\left(1,2,1,2\right)$, while the latter states have large weights in $\left(1,2,2,1\right)$ or $\left(2,1,1,2\right)$.

Additionally to \eref{eq:Coul2}, there are also the single-qubit interactions
\begin{align}
\mathcal{C}_{1Q}
=&
X_{\mathcal{C}}^{\text{L}}
\sigma_x^{\text{L}}
+
X_{\mathcal{C}}^{\text{R}}
\sigma_x^{\text{R}},
\label{eq:Coul3}
\end{align}
with $X_{\mathcal{C}}^{\text{L}}=
-\frac{\kappa}{4}
\sin^2\left(\theta^{\text{L}}_*/2\right)
\sin^2\left(\theta^{\text{R}}_*/2\right)$ and 
$X_{\mathcal{C}}^{\text{R}}=
\frac{\kappa}{4}
[1+\cos^2\left(\theta^{\text{L}}_*/2\right)]
\sin^2\left(\theta^{\text{R}}_*/2\right)$.

\subsection{\label{sec:Exc}
Exchange Coupling}

\begin{figure}
\includegraphics[width=0.49\textwidth]{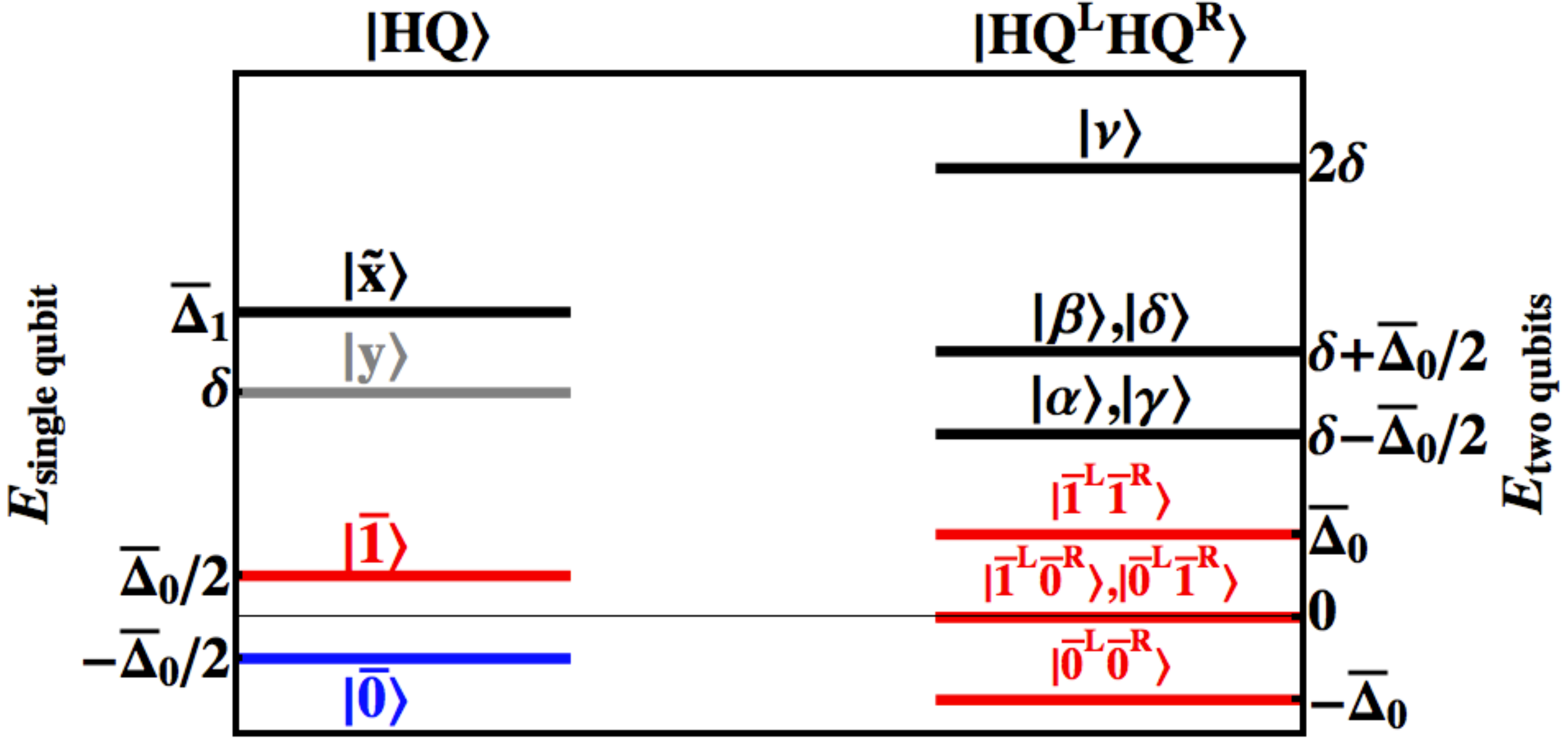}
\caption{\label{fig:03}
States with the lowest energies for a HQ and for the combined system of two identical HQs in the absence of inter-qubit exchange and Coulomb interactions. Each HQ is operated at $\epsilon_*=\epsilon_*^{\text{L}}=\epsilon_*^{\text{R}}$. The states of a HQ are shown on the left (cf. \fref{fig:02}); the qubit states are labeled by $\ket{\overline{1}}$ and $\ket{\overline{0}}$. The leakage states are the excited, hybridized state between the $\left(2,1\right)$ and $\left(1,2\right)$ configurations $\ket{\widetilde{\text{x}}}$ and the $S=3/2$, $s_z=1/2$ triplet state $\ket{\text{y}}$. The two-qubit states are shown on the right. The computational subspace
$\left\{\ket{\overline{1}^{\text{L}}\overline{1}^{\text{R}}}, 
\ket{\overline{1}^{\text{L}}\overline{0}^{\text{R}}},
\ket{\overline{0}^{\text{L}}\overline{1}^{\text{R}}},
\ket{\overline{0}^{\text{L}}\overline{0}^{\text{R}}}\right\}$ is well separated in energy from all the leakage states $\ket{\alpha}$, $\ket{\beta}$, $\ket{\gamma}$, $\ket{\delta}$, and $\ket{\nu}$ that are defined in \tref{tab:01}.
}
\end{figure}

Encoded qubits in the multielectron regime have a large number of spin states that are not part of the computational subspace.\cite{kempe2001} As a consequence, the time evolution out of the computational subspace, which is called leakage, must be considered.\cite{taylor2005,mehl2015-2} Similar to TQDs, inter-qubit tunnel couplings between HQs can cause leakage because the subspaces of different local spin quantum numbers are coupled.\cite{divincenzo2000,fong2011} It will be shown that the system of two HQs at their sweet spots have large energy separations between leakage and computational states, which suppresses leakage events (cf. similar approaches for TQDs in \mrcite{weinstein2005}{doherty2013}).

\begin{table*}
\caption{\label{tab:01}
Relevant two-qubit states with total $s_z=1$ for two HQs. All the states of a three-electron DQD with $S=1/2$, $s_z=\pm1/2$ have contributions in the $\left(2,1\right)$ and $\left(1,2\right)$ configurations. The computational subspace
$\left\{\ket{\overline{1}^{\text{L}}\overline{1}^{\text{R}}},
\ket{\overline{1}^{\text{L}}\overline{0}^{\text{R}}},
\ket{\overline{0}^{\text{L}}\overline{1}^{\text{R}}},
\ket{\overline{0}^{\text{L}}\overline{0}^{\text{R}}}\right\}$ is energetically separated from all the leakage states. I use the abbreviations $\ket{\overline{v}_1}$ and $\ket{\overline{v}_0}$ for the basis states in $S=1/2$, $s_z=-1/2$ that are obtained from $\ket{\overline{1}}$ and $\ket{\overline{0}}$ by flipping all the spins. The states $\ket{\alpha}$, $\ket{\beta}$, $\ket{\gamma}$, $\ket{\delta}$, and $\ket{\nu}$ label the leakage states with the energies 
$\frac{\overline{\Delta}_0^{\text{L}}}{2}+\delta^{\text{R}}$,
$-\frac{\overline{\Delta}_0^{\text{L}}}{2}+\delta^{\text{R}}$,
$\delta^{\text{L}}+\frac{\overline{\Delta}_0^{\text{R}}}{2}$,
$\delta^{\text{L}}-\frac{\overline{\Delta}_0^{\text{R}}}{2}$, and 
$\delta^{\text{L}}+\delta^{\text{R}}$.
}
\begin{tabular}{llll}
\hline
\hline
\multicolumn{1}{c}{state}
& \multicolumn{1}{c}{$\left(\text{S}^{\text{L}},s_z^{\text{L}}\right)$}
& \multicolumn{1}{c}{$\left(\text{S}^{\text{R}},s_z^{\text{R}}\right)$}
& \multicolumn{1}{c}{energy}\\
$\ket{\overline{1}^{\text{L}}\overline{1}^{\text{R}}}$
& 
\rdelim\}{6}{4mm}[$\left(\frac{1}{2},\frac{1}{2}\right)$] &
\rdelim\}{6}{4mm}[$\left(\frac{1}{2},\frac{1}{2}\right)$]
&$\left(\overline{\Delta}_0^{\text{L}}+\overline{\Delta}_0^{\text{R}}\right)/2$\\
$\ket{\overline{1}^{\text{L}}\overline{0}^{\text{R}}}$
& &
&$\left(\overline{\Delta}_0^{\text{L}}-\overline{\Delta}_0^{\text{R}}\right)/2$\\
$\ket{\overline{0}^{\text{L}}\overline{1}^{\text{R}}}$
& &
&$\left(-\overline{\Delta}_0^{\text{L}}+\overline{\Delta}_0^{\text{R}}\right)/2$\\
$\ket{\overline{0}^{\text{L}}\overline{0}^{\text{R}}}$
& &
&$-\left(\overline{\Delta}_0^{\text{L}}+\overline{\Delta}_0^{\text{R}}\right)/2$\\
\\
$\ket{\alpha_1}=
\ket{\overline{1}^{\text{L}}}
\left[\sqrt{\frac{1}{3}}\ket{\downarrow \text{T}_+}+\sqrt{\frac{2}{3}}\ket{\uparrow \text{T}_0}\right]$
& $\left(\frac{1}{2},\frac{1}{2}\right)$ & $\left(\frac{3}{2},\frac{1}{2}\right)$
&\rdelim\}{2}{4mm}[$\frac{\overline{\Delta}_0^{\text{L}}}{2}+\delta^{\text{R}}$]\\
$\ket{\alpha_2}=
\ket{\overline{v}_1^{\text{L}}}
\ket{\uparrow \text{T}_+}$
& $\left(\frac{1}{2},-\frac{1}{2}\right)$ & $\left(\frac{3}{2},\frac{3}{2}\right)$\\
\\
$\ket{\beta_1}=
\ket{\overline{0}^{\text{L}}}
\left[\sqrt{\frac{1}{3}}\ket{\downarrow \text{T}_+}+\sqrt{\frac{2}{3}}\ket{\uparrow \text{T}_0}\right]$
& $\left(\frac{1}{2},\frac{1}{2}\right)$ & $\left(\frac{3}{2},\frac{1}{2}\right)$
&\rdelim\}{2}{4mm}[$-\frac{\overline{\Delta}_0^{\text{L}}}{2}+\delta^{\text{R}}$]\\
$\ket{\beta_2}=
\ket{\overline{v}_0^{\text{L}}}
\ket{\uparrow \text{T}_+}$
& $\left(\frac{1}{2},-\frac{1}{2}\right)$ & $\left(\frac{3}{2},\frac{3}{2}\right)$\\
\\
$\ket{\gamma_1}=
\left[\sqrt{\frac{1}{3}}\ket{\downarrow \text{T}_+}+\sqrt{\frac{2}{3}}\ket{\uparrow \text{T}_0}\right]
\ket{\overline{1}^{\text{R}}}$
& $\left(\frac{3}{2},\frac{1}{2}\right)$ & $\left(\frac{1}{2},\frac{1}{2}\right)$
&\rdelim\}{2}{4mm}[$\delta^{\text{L}}+\frac{\overline{\Delta}_0^{\text{R}}}{2}$]\\
$\ket{\gamma_2}=
\ket{\uparrow \text{T}_+}
\ket{\overline{v}_1^{\text{R}}}$
& $\left(\frac{3}{2},\frac{3}{2}\right)$ & $\left(\frac{1}{2},-\frac{1}{2}\right)$\\
\\
$\ket{\delta_1}=
\left[\sqrt{\frac{1}{3}}\ket{\downarrow \text{T}_+}+\sqrt{\frac{2}{3}}\ket{\uparrow \text{T}_0}\right]
\ket{\overline{0}^{\text{R}}}$
& $\left(\frac{3}{2},\frac{1}{2}\right)$ & $\left(\frac{1}{2},\frac{1}{2}\right)$
&\rdelim\}{2}{4mm}[$\delta^{\text{L}}-\frac{\overline{\Delta}_0^{\text{R}}}{2}$]\\
$\ket{\delta_2}=
\ket{\uparrow \text{T}_+}
\ket{\overline{v}_0^{\text{R}}}$
& $\left(\frac{3}{2},\frac{3}{2}\right)$ & $\left(\frac{1}{2},-\frac{1}{2}\right)$\\
\\
$\ket{\nu_1}=
\left[\sqrt{\frac{1}{3}}\ket{\downarrow \text{T}_+}+\sqrt{\frac{2}{3}}\ket{\uparrow \text{T}_0}\right]
\left[\sqrt{\frac{1}{3}}\ket{\downarrow \text{T}_+}+\sqrt{\frac{2}{3}}\ket{\uparrow \text{T}_0}\right]$
& $\left(\frac{3}{2},\frac{1}{2}\right)$ & $\left(\frac{3}{2},\frac{1}{2}\right)$
&\rdelim\}{4}{4mm}[$\delta^{\text{L}}+\delta^{\text{R}}$]\\
$\ket{\nu_2}=
\left[\sqrt{\frac{1}{3}}\ket{\uparrow \text{T}_-}+\sqrt{\frac{2}{3}}\ket{\downarrow \text{T}_0}\right]
\ket{\uparrow \text{T}_+}$
& $\left(\frac{3}{2},-\frac{1}{2}\right)$ & $\left(\frac{3}{2},\frac{3}{2}\right)$\\
$\ket{\nu_3}=
\ket{\uparrow \text{T}_+}
\left[\sqrt{\frac{1}{3}}\ket{\uparrow \text{T}_-}+\sqrt{\frac{2}{3}}\ket{\downarrow \text{T}_0}\right]$
& $\left(\frac{3}{2},\frac{3}{2}\right)$ & $\left(\frac{3}{2},-\frac{1}{2}\right)$\\
\hline
\hline
\end{tabular}
\end{table*}

In the six-electron configuration, only states with total $s_z=1$ are considered because the spin directions are conserved during tunneling events. \tref{tab:01} summarizes the spin states with the lowest energies. The computational subspace
$\left\{\ket{\overline{1}^{\text{L}}\overline{1}^{\text{R}}},
\ket{\overline{1}^{\text{L}}\overline{0}^{\text{R}}},
\ket{\overline{0}^{\text{L}}\overline{1}^{\text{R}}},
\ket{\overline{0}^{\text{L}}\overline{0}^{\text{R}}}\right\}$ and the leakage states (called $\ket{\alpha}$, $\ket{\beta}$, $\ket{\gamma}$, $\ket{\delta}$, and $\ket{\nu}$) are considered. Only the states with $S=1/2$ are able to couple the $\left(2,1\right)$ and $\left(1,2\right)$ configurations at QD$_{1}^{\text{L}}$ and QD$_{2}^{\text{L}}$ or at QD$_{1}^{\text{R}}$ and QD$_{2}^{\text{R}}$. These states have lower energies than the state with $S=3/2$ of the same $s_z$ (cf. \fref{fig:02}: $\ket{\overline{1}}$ and $\ket{\overline{0}}$ have $S=1/2$, $\ket{\text{y}}$ has $S=3/2$). As a consequence, all the computational states have lower energies than the leakage states (unless $\delta^{\text{L}},\delta^{\text{R}}\lesssim\frac{\overline{\Delta}_0^{\text{L}}}{2},\frac{\overline{\Delta}_0^{\text{R}}}{2}$). \fref{fig:03} shows the energies of the single-qubit states and the leakage states for two identical HQs.

A tunnel coupling between QD$_2^{\text{L}}$ and QD$_1^{\text{R}}$ couples the $\left(2,1\right)$ and $\left(1,2\right)$ configurations at these QDs. Similar to the single-qubit interactions in \sref{sec:Single}, I phenomenologically introduce a state coupling between QD$_{2}^{\text{L}}$ and QD$_{1}^{\text{R}}$:
\begin{align}
\label{eq:Pert}
\mathcal{H}^\prime=
&-t_1
\ket{\uparrow \text{S}}\left(\sqrt{\frac{2}{3}}\bra{\text{T}_+\downarrow}-\sqrt{\frac{1}{3}}\bra{\text{T}_0\uparrow}\right)\\
&+t_0
\op{\uparrow \text{S}}{\text{S} \uparrow}
+\text{h.c.}
\nonumber
\end{align}
Note that only the states with $S=1/2$, $s_z=1/2$ at QD$_{2}^{\text{L}}$ and QD$_{1}^{\text{R}}$ are coupled in \eref{eq:Pert}. All triplets at QD$_1^{\text{R}}$ are neglected because these states have much higher energies.

The setup can be tuned towards the $\left(1,1,2,2\right)$ configuration. Only states with a singlet at QD$_1^{\text{R}}$ are considered, giving the four states
$\{\ket{\uparrow\uparrow \text{S S}}, \ket{\uparrow\uparrow \text{S T}_0}, \ket{\uparrow\downarrow \text{S T}_+}, 
\ket{\downarrow\uparrow \text{S T}_+}\}$ with $s_z=1$.  If the $\left(1,1,2,2\right)$ spin configuration is only virtually excited, then $\mathcal{H}^\prime$ can be eliminated using Schrieffer-Wolff perturbation theory.\cite{schrieffer1966,winkler2010,bravyi2011} Only the effective interactions on the qubit subspace are summarized in the following; \aref{app:Exc} gives a detailed summary of their derivations.

There is an effective two-qubit interaction
\begin{align}
\mathcal{E}_{2Q}=
\mathcal{X}_{\mathcal{E}}\sigma_x^{\text{L}}\sigma_x^{\text{R}}+
\mathcal{Z}_{\mathcal{E}}\sigma_z^{\text{L}}\sigma_x^{\text{R}},
\label{eq:Ex1}
\end{align}
with 
$
\mathcal{X}_{\mathcal{E}}=-
\left[\frac{1-c_{\text{R}}^2}{4}\right]J_0+
\left[\frac{c_{\text{L}}^2\left(3-11c_{\text{R}}^2\right)}{108}\right]J_1
$ and 
$
\mathcal{Z}_{\mathcal{E}}=-
\left[\frac{c_{\text{L}}\left(1-c_{\text{R}}^2\right)}{6}\right]\sqrt{J_0J_1}$. $J_0=\frac{t_0^2}{E_{\left(1,1,2,2\right)}-E_{\left(1,2,1,2\right)}}$ and 
$J_1=\frac{t_1^2}{E_{\left(1,1,2,2\right)}-E_{\left(1,2,1,2\right)}}$ are the effective coupling constants that are determined by the energy difference $E_{\left(1,1,2,2\right)}-E_{\left(1,2,1,2\right)}$ between $\left(1,1,2,2\right)$ and $\left(1,2,1,2\right)$. $c_{\text{L}}$ [$c_{\text{R}}$] is the abbreviation for $\cos\left(\theta_*^{\text{L}}/2\right)$ [$\cos\left(\theta_*^{\text{R}}/2\right)$]. Note that there is additionally a single-qubit Hamiltonian
\begin{align}
\mathcal{E}_{1Q}=&
Z^{\text{L}}_{\mathcal{E}}\sigma_z^{\text{L}}+
X^{\text{L}}_{\mathcal{E}}\sigma_x^{\text{L}}+
X^{\text{R}}_{\mathcal{E}}\sigma_x^{\text{R}},
\label{eq:Ex2}
\end{align}
with
$Z^{\text{L}}_{\mathcal{E}}=
\left[\frac{c_{\text{L}}\left(1+c_{\text{R}}^2\right)}{6}\right]
\sqrt{J_0J_1}$,
$X^{\text{L}}_{\mathcal{E}}=
\left[\frac{1+c_{\text{R}}^2}{4}\right]J_0-
\left[\frac{c_{\text{L}}^2\left(3+11c_{\text{R}}^2\right)}{108}\right]J_1$,
and 
$X^{\text{R}}_{\mathcal{E}}=
\left[\frac{1-c_{\text{R}}^2}{4}\right]J_0+
\left[\frac{c_{\text{L}}^2\left(3-11c_{\text{R}}^2\right)}{108}\right]J_1$.

Besides \eref{eq:Ex1} and \eref{eq:Ex2}, the effective Hamiltonian also contains contributions for the leakage states and between leakage and computational states. All these contributions will be included in the numerical calculations that contain inter-qubit exchange interactions. Assuming large energy separations between the computational and leakage states, these interactions introduce minor effects for the qubit's time evolution, and they can be neglected.

\section{\label{sec:UnivCont}
Universal Qubit Control}

I discuss gate operations for the parameters (A) $\overline{\Delta}_0^{\text{L}}\approx\overline{\Delta}_0^{\text{R}}$ and (B) $\overline{\Delta}_0^{\text{L}}\gg\overline{\Delta}_0^{\text{R}}$. Note that the HQs are always operated at their sweet spots $\epsilon_*^{\text{L}}$ and $\epsilon_*^{\text{R}}$.

\subsection{\label{sec:UnivA}
Nearly identical qubits $\overline{\Delta}_0^{\text{L}}\approx\overline{\Delta}_0^{\text{R}}$}

\begin{figure}
\includegraphics[width=0.49\textwidth]{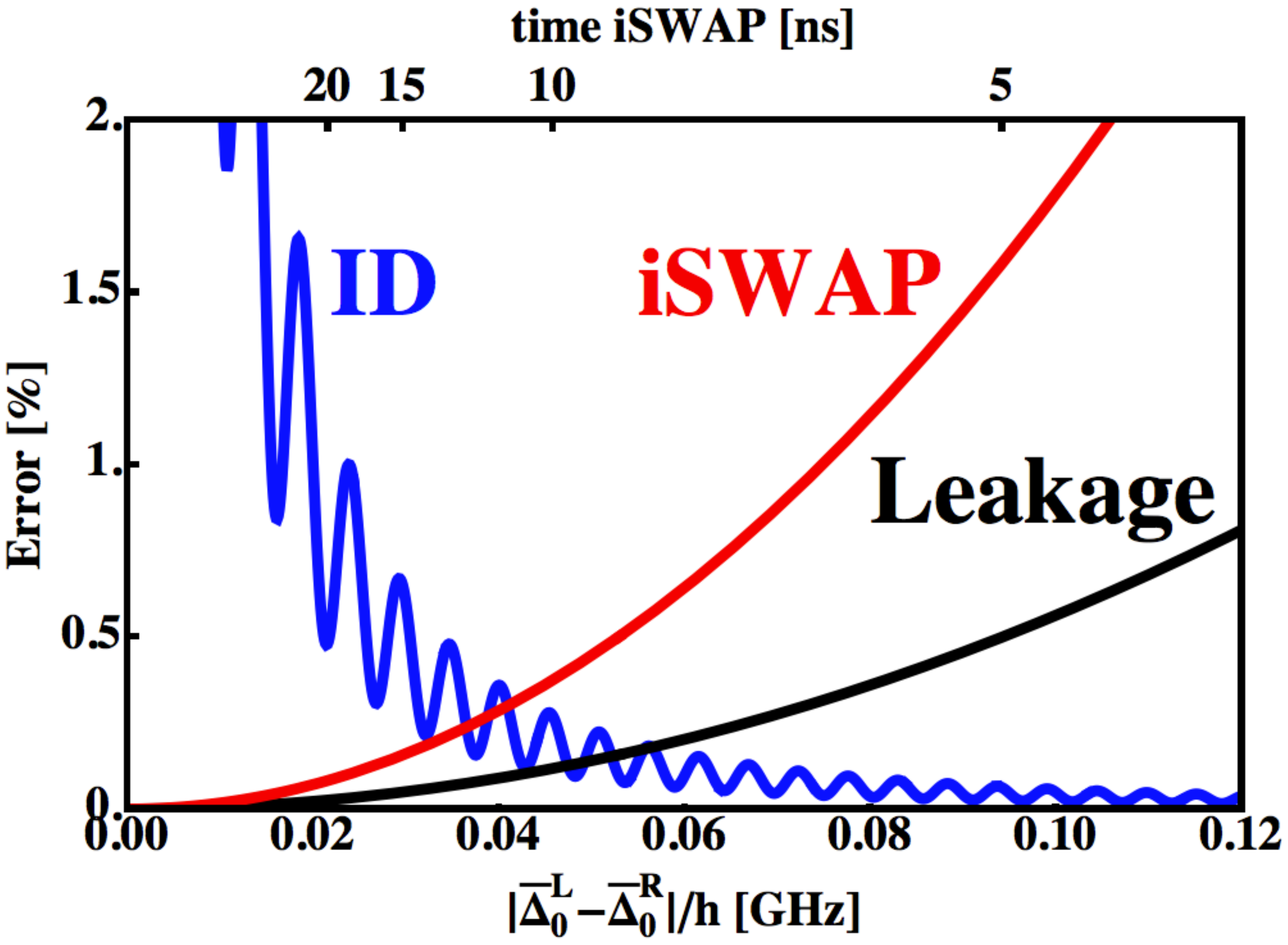}
\caption{\label{fig:04}
Gate operations for two HQs with similar eigenfrequencies $\overline{\Delta}_0^{\text{L}}\approx\overline{\Delta}_0^{\text{R}}$ that use the static Coulomb interaction according to \eref{eq:Coul2}, with $\mathcal{X}_{\mathcal{C}}/h\approx 1.3~\text{MHz}$, and the tunable exchange interactions from \eref{eq:Ex1}-\eqref{eq:Ex2}. A small detuning $\overline{\Delta}_0^{\text{R}}-\overline{\Delta}_0^{\text{L}}\gg\mathcal{X}_{\mathcal{C}}$ is sufficient to neglect the interactions between HQ\tu{L} and HQ\tu{R} such that the qubits evolve independently. ID shows the infidelity, according to \eref{eq:Fidelity}, of a static time evolution with the given parameters and $\mathcal{X}=\mathcal{X}_{\mathcal{C}}$, which is compared to the ideal time evolution with $\mathcal{X}=0$ after the time $t=h/(4\mathcal{X}_{\mathcal{C}})$. Adding a small tunnel coupling between QD$_2^{\text{L}}$ and QD$_1^{\text{R}}$ and tuning the spin configuration towards $\left(1,1,2,2\right)$ brings HQ\tu{L} into resonance with HQ\tu{R} (cf. \eref{eq:Ex2}, $Z^{\text{L}}_{\mathcal{E}}=\overline{\Delta}_0^{\text{R}}-\overline{\Delta}_0^{\text{L}}$). The HQs entangle after $t=h/(4\mathcal{X})$ for $\mathcal{X}=\mathcal{X}_{\mathcal{C}}+\mathcal{X}_{\mathcal{E}}$. iSWAP shows the infidelity, according to \eref{eq:Fidelity}, of the entangling operation with the given parameters, which is compared to an ideal entangling operation according to \eref{eq:HamA}. The qubits can be operated independently for $\abs{\overline{\Delta}_0^{\text{R}}-\overline{\Delta}_0^{\text{L}}}/h>20~\text{MHz}$. When the inter-qubit exchange interaction increases, HQ\tu{L} and HQ\tu{R} are brought into resonance to construct high-fidelity entangling operations. Note that also leakage errors increase with the inter-qubit exchange interactions, as discussed in the text.
}
\end{figure}

I consider the Hamiltonian:
\begin{align}
\mathcal{H}_{\text{A}}=
\frac{
\overline{\Delta}_0^{\text{L}}+Z^{\text{L}}
}{2}\sigma_z^{\text{L}}+
\frac{
\overline{\Delta}_0^{\text{R}}
}{2}\sigma_z^{\text{R}}+
\mathcal{X}\sigma_x^{\text{L}}\sigma_x^{\text{R}}.
\label{eq:HamA}
\end{align}
If HQ\tu{L} and HQ\tu{R} are in resonance, $Z^{\text{L}}=\overline{\Delta}_0^{\text{R}}-\overline{\Delta}_0^{\text{L}}$, then HQ\tu{L} and HQ\tu{R} entangle under static time evolutions. In this case, the two-qubit interaction is
\begin{align}
\mathcal{H}^{\text{rwa}}_{\text{A}}=
\frac{\mathcal{X}}{2}
\left(\sigma_x^{\text{L}}\sigma_x^{\text{R}}+\sigma_y^{\text{L}}\sigma_y^{\text{R}}\right)
\end{align}
in the rotating frame with $\frac{\overline{\Delta}_0^{\text{R}}}{2}\left(
\sigma_z^{\text{L}}+\sigma_z^{\text{R}}\right)$ for $\overline{\Delta}_0^{\text{R}}\gg \mathcal{X}$. After the time $t=\frac{h}{4\mathcal{X}}$, the $\text{iSWAP}$ gate is created, which is maximally entangling.\footnote{The CPHASE gate can be obtained using the sequence
\begin{align*}
e^{-i\frac{\pi}{4}\sigma_z^{\text{R}}}&
e^{-i\frac{\pi}{4}\sigma_x^{\text{R}}}
\text{iSWAP}
e^{i\frac{\pi}{4}\sigma_z^{\text{L}}}
\text{iSWAP}
e^{-i\frac{\pi}{4}\sigma_z^{\text{L}}}
e^{-i\frac{\pi}{4}\sigma_x^{\text{R}}}\\&
=e^{-i\frac{3\pi}{4}}\text{CPHASE},
\end{align*}
such that two iSWAPs construct one CPHASE.
}\textsuperscript{,}\cite{schuch2003} On the other hand, the two-qubit interaction in \eref{eq:HamA} can be neglected for $\overline{\Delta}_0^{\text{L}}+Z^{\text{L}}\neq\overline{\Delta}_0^{\text{R}}\gg\mathcal{X}$ and all single-qubit gates can be realized with the methods that were described in \sref{sec:Single}.

Altogether, universal qubit control requires rapid control over $Z^{\text{L}}$. For $Z^{\text{L}}=0$, all the single-qubit gates can be realized with resonant drivings of the qubits at their eigenfrequencies. For $Z^{\text{L}}=\overline{\Delta}_0^{\text{R}}-\overline{\Delta}_0^{\text{L}}$, the HQs entangle under static time evolutions.

In the configurations of \fref{fig:01}, $\mathcal{X}$ in \eref{eq:HamA} has contributions from the capacitive couplings between the charge configurations $\mathcal{X}_{\mathcal{C}}$ (cf. \sref{sec:Cap}) and from the exchange interaction $\mathcal{X}_{\mathcal{E}}$ between QD$_2^{\text{L}}$ and QD$_1^{\text{R}}$ (cf. \sref{sec:Exc}): $\mathcal{X}=\mathcal{X}_{\mathcal{C}}+\mathcal{X}_{\mathcal{E}}$. The exchange interaction causes also a frequency shift of HQ\tu{L} [$Z^{\text{L}}_{\mathcal{E}}$ in \eref{eq:Ex2}] which can be controlled electrically. Note that the Coulomb interaction remains constant during the qubit manipulations because the HQs are always operated at their sweet spots. The single-qubit interactions from \eref{eq:Coul3} are neglected because they only modify the positions of the anticrossings $\epsilon_*^{\text{L}}$ and $\epsilon_*^{\text{R}}$.

$\overline{\Delta}_0^{\text{L}}$ and $\overline{\Delta}_0^{\text{R}}$ should have similar magnitudes such that only small inter-qubit exchange interactions are needed to bring HQ\tu{L} and HQ\tu{R} into resonance. Still, $\overline{\Delta}_0^{\text{L}}$ and $\overline{\Delta}_0^{\text{R}}$ are sufficiently distinct for independent single-qubit control. The capacitive coupling $\mathcal{X}_C$ between the HQs must remain much smaller than in the naive approximations in \sref{sec:Cap}.
\fref{fig:04} shows simulations with $\kappa=0.04~\mu\text{eV}$ and all the parameters defined earlier. For the Coulomb interactions, only the two-qubit interaction $\mathcal{X}_C$ is taken into account according to \eref{eq:Coul2}. The exchange interaction contains all the contributions from \eref{eq:Ex1} and \eref{eq:Ex2}.

If the exchange interactions between HQ\tu{L} and HQ\tu{R} [$Z^{\text{L}}_{\mathcal{E}}$ in \eref{eq:Ex2} and $Z^{\text{L}}$ in \eref{eq:HamA}] cancel the detuning between the HQs, then the entangling gate is realized. Simultaneously with the single-qubit energy shift, also $X^{\text{L}}_{\mathcal{E}}$ and $X^{\text{R}}_{\mathcal{E}}$ from \eref{eq:Ex2} increase. These interactions cause systematic gate errors. \fref{fig:04} shows that the gate errors of the iSWAP below $1\%$ can be realized for $\abs{\overline{\Delta}_0^{\text{R}}-\overline{\Delta}_0^{\text{L}}}/h<80~\text{MHz}$. Leakage errors are also caused by the exchange couplings between HQ\tu{L} and HQ\tu{R}, and these leakage events significantly contribute to the gate infidelities. \fref{fig:05} extracts the leakage errors during an entangling gate with finite tunnel couplings between HQ\tu{L} and HQ\tu{R}. The leakage probability is extracted from the time evolution $\mathcal{U}$ by taking the norm of the matrix $\abs{\mathcal{U}_{\mathcal{P}_{\mathcal{C}}\mathcal{P}_{\mathcal{L}}}}^2$ between the states from the computational subspace $\mathcal{P}_{\mathcal{C}}$ and the leakage subspace $\mathcal{P}_{\mathcal{L}}$.

If $Z^{\text{L}}$ is reduced, then HQ\tu{L} and HQ\tu{R} decouple. \fref{fig:04} compares the time evolution of \eref{eq:HamA} with the time evolution of
\begin{align}
\mathcal{H}_{\text{A}}^{\text{ideal}}=\frac{
\overline{\Delta}_0^{\text{L}}
}{2}\sigma_z^{\text{L}}+
\frac{
\overline{\Delta}_0^{\text{R}}
}{2}\sigma_z^{\text{R}}.
\end{align}
The differences rise significantly if $\abs{\overline{\Delta}^{\text{L}}_0-\overline{\Delta}^{\text{R}}_0}$ decreases; but the infidelities stay below 1\% for $\abs{\overline{\Delta}^{\text{L}}_0-\overline{\Delta}^{\text{R}}_0}/h>20~\text{MHz}$.

\subsection{\label{sec:UnivB}
Distinct qubits $\overline{\Delta}_0^{\text{L}}\gg\overline{\Delta}_0^{\text{R}}$}

\begin{figure}
\includegraphics[width=0.49\textwidth]{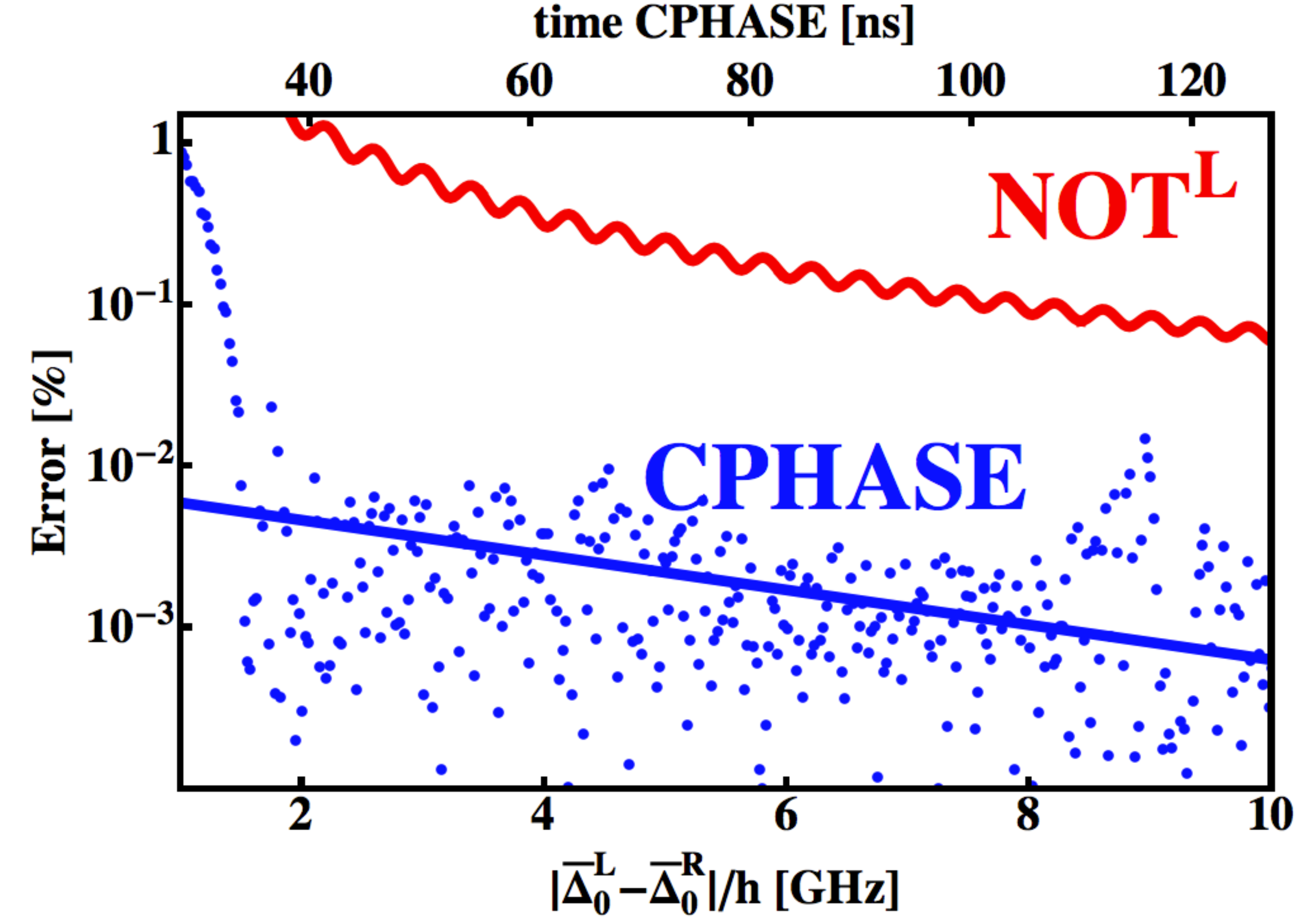}
\caption{\label{fig:05}
Resonant manipulations of two HQs with large detunings between their eigenfrequencies $\overline{\Delta}_0^{\text{L}}/h$ and $\overline{\Delta}_0^{\text{R}}/h$ of a few GHz. The qubits interact via the Coulomb interaction, according to \eref{eq:Coul2}, with $\mathcal{X}_{\mathcal{C}}/h=0.13~\text{GHz}$. A resonant drive of HQ\tu{L} at its eigenfrequency realizes a state inversion of HQ\tu{L}, while HQ\tu{R} evolves trivially. The NOT$^{\text{L}}$ gate is realized after the time $t=h/(4\mathcal{A})$ for a drive $\mathcal{A}\cos\left(\overline{\Delta}^{\text{L}}_0t/h\right)$ with $\mathcal{A}/h=0.1~\text{GHz}$. The time evolution is compered to an ideal NOT$^{\text{L}}$ with $\mathcal{X}_{\mathcal{C}}=0$, showing that the gate infidelity, according to \eref{eq:Fidelity}, is lower than 1\%. A similar drive of HQ\tu{L} with the frequency $\overline{\Delta}_0^{\text{R}}/h$ entangles HQ\tu{L} and HQ\tu{R}. The resulting gate operation is equivalent to a CPHASE gate up to single-qubit gates. In the simulations, the gate times of the entangling gates are optimized numerically to cancel high-frequency fluctuations. The gate errors are quantified by the deviations of the Makhlin invariants from their ideal values ($\abs{G_1}+\abs{G_2-1}$).\cite{makhlin2002} $\abs{\overline{\Delta}_0^{\text{L}}-\overline{\Delta}_0^{\text{R}}}/h>2~\text{GHz}$ permits high-fidelity CPHASE gates.
}
\end{figure}

Highly detuned HQs (e.g. $\overline{\Delta}_0^{\text{L}}\gg\overline{\Delta}_0^{\text{R}}$) can be operated exclusively with microwave signals. The $\sigma_x^{\text{L}}\sigma_x^{\text{R}}$ interaction between HQ\tu{L} and HQ\tu{R} can be neglected without drivings, and if both qubits are driven with their own resonance frequencies. In the cross-resonance protocol, one qubit is driven at the eigenfrequency of the other qubit \cite{paraoanu2006,rigetti2010,chow2011,chow2012}, e.g., 
\begin{align}
\mathcal{H}_{\text{B}}=
\frac{
\overline{\Delta}_0^{\text{L}}
}{2}\sigma_z^{\text{L}}+
\mathcal{A}\cos\left(2\pi\frac{\overline{\Delta}_0^{\text{R}}t}{h}\right)\sigma_x^{\text{L}}+
\frac{
\overline{\Delta}_0^{\text{R}}
}{2}\sigma_z^{\text{R}}+
\mathcal{X}\sigma_x^{\text{L}}\sigma_x^{\text{R}}.
\label{eq:Cross1}
\end{align}
Transforming \eref{eq:Cross1} to the rotating frame with $\frac{\overline{\Delta}_0^{\text{R}}}{2}\left(\sigma_z^{\text{L}}+\sigma_z^{\text{R}}\right)$ gives after a rotating wave approximation
\begin{align}
\mathcal{H}^{\text{rwa}}_{\text{B}}=
\frac{
\overline{\Delta}_0^{\text{L}}-\overline{\Delta}_0^{\text{R}}
}{2}\sigma_z^{\text{L}}+
\frac{\mathcal{A}}{2}\sigma_x^{\text{L}}+
\frac{\mathcal{X}}{2}\left(
\sigma_x^{\text{L}}\sigma_x^{\text{R}}+
\sigma_y^{\text{L}}\sigma_y^{\text{R}}
\right).
\label{eq:Cross2}
\end{align}
Another rotating wave approximation gives the effective interaction
\begin{align}
\label{eq:RWA2}
\mathcal{H}^{\text{rwa}_2}_{\text{B}}=&
\frac{\mathcal{X}}{2}\frac{\mathcal{A}}{\sqrt{\left(\overline{\Delta}_0^{\text{L}}-\overline{\Delta}_0^{\text{R}}\right)^2+\mathcal{A}^2}}
\\\nonumber&\times
\left[\cos\left(\vartheta\right)\sigma_z^{\text{L}}+\sin\left(\vartheta\right)\sigma_x^{\text{L}}\right]\sigma_x^{\text{R}}
\end{align}
in the rotating frame with $\frac{
\overline{\Delta}_0^{\text{L}}-\overline{\Delta}_0^{\text{R}}
}{2}\sigma_z^{\text{L}}+
\frac{\mathcal{A}}{2}\sigma_x^{\text{L}}$. I use the abbreviations
$\cos\left(\vartheta\right)=\frac{
\overline{\Delta}_0^{\text{L}}-\overline{\Delta}_0^{\text{R}}
}{\sqrt{\left(\overline{\Delta}_0^{\text{L}}-\overline{\Delta}_0^{\text{R}}\right)^2+\mathcal{A}^2}}$ and $\sin\left(\vartheta\right)=\frac{
\mathcal{A}
}{\sqrt{\left(\overline{\Delta}_0^{\text{L}}-\overline{\Delta}_0^{\text{R}}\right)^2+\mathcal{A}^2}}$ in \eref{eq:RWA2}. A drive for the time $
t=
\frac{\mathcal{X}}{16}
\frac{\mathcal{A}}{\sqrt{\left(\overline{\Delta}_0^{\text{L}}-\overline{\Delta}_0^{\text{R}}\right)^2+\mathcal{A}^2}}$ entangles the HQs because the time evolution of \eref{eq:RWA2} is equivalent to a CPHASE gate.\footnote{\eref{eq:RWA2} is in a rotated basis compared to \protect{$\sigma_z^{\text{L}}\sigma_z^{\text{R}}$}, but it generates the equivalent entangling gate. It is well known that \protect{$\sigma_z^{\text{L}}\sigma_z^{\text{R}}$} is maximally entangling with
\begin{align*}
e^{-i\frac{\pi}{4}\sigma_z^{\text{L}}\sigma_z^{\text{R}}}=
e^{i\frac{\pi}{4}}
e^{-i\frac{5\pi}{4}\sigma_z^{\text{L}}}
e^{-i\frac{5\pi}{4}\sigma_z^{\text{R}}}
\text{CPHASE}.
\end{align*}
CPHASE has the Makhlin invariants\cite{makhlin2002} $G_1=0$ and $G_2=1$.
}

This entangling operation only requires the $\sigma_x^{\text{L}}\sigma_x^{\text{R}}$ interaction, which can be obtained from the Coulomb interaction between the HQs [see \eref{eq:Coul2}]. The numerical simulations use $\kappa=4~\mu\text{eV}$, which is smaller than the suggested value from the naive estimates in \sref{sec:Cap}. Because the inter-qubit exchange interactions are not needed, which would require QDs in close vicinity, one can reduce $\kappa$ in \eref{eq:Coul1} by sufficiently separating the DQDs. In this case, there are no leakage errors (see \sref{sec:Exc}) because the electron transfer between the QDs is forbidden. I simulate the time evolutions according to \eref{eq:Cross1} without any rotating wave approximations. Besides the dominant time evolution from \eref{eq:RWA2}, there are rapidly oscillating terms that were neglected in the rotating wave approximations in \eref{eq:Cross2} and \eref{eq:RWA2}.

\fref{fig:05} shows the errors of entangling operations for $\mathcal{X}=0.55~\mu\text{eV}$, driving amplitudes $\mathcal{A}/h=0.1~\text{GHz}$, and the HQ parameters defined earlier.  To realize high-fidelity CPHASE operations, the deviations of the Makhlin invariants from $G_1=0$ and $G_2=1$ are minimized. This routine cancels artifacts that are neglected in the rotating wave approximations of \eref{eq:Cross1} and \eref{eq:Cross2}. If the HQs are detuned by several GHz, then high-fidelity entangling operations can be realized. \fref{fig:05} also shows that high-fidelity single-qubit operations can be realized in the same setup if the HQs' eigenfrequencies are detuned by a few GHz.

\section{\label{sec:Noise}
Noise Discussion}

This section shows that noise from fluctuating magnetic and electric fields only causes small errors for HQs. In the following, a HQ is analyzed in the eigenbasis $\left\{\ket{\overline{1}},\ket{\overline{0}}\right\}$ according to \eref{eq:Ham3} because it is operated at its sweet spot $\epsilon_*$.

\subsection{Hyperfine Interactions}
Fluctuating local magnetic fields were identified as a natural problem for spin qubits.\cite{burkard1999} The nuclear magnetic fields of the host's nuclei couples via the contact hyperfine interaction to the spin of an electron\cite{hanson2007-2,taylor2007,coish2009}. For localized electrons, the contact hyperfine interaction can be described by slowly-fluctuating local magnetic fields at the QDs: $\delta \bm{B}^{\text{QD}_1}$ and $\delta \bm{B}^{\text{QD}_2}$. 

For the HQ, a noise term
$\mathcal{H}^{\delta \bm{B}}=\frac{g\mu_B}{2}(
\delta \bm{B}^{\text{QD}_1}\cdot\bm{\sigma}^{\text{QD}_1}+\delta \bm{B}^{\text{QD}_2}\cdot\bm{\sigma}^{\text{QD}_2})$
describes the fluctuating magnetic fields, where
$\bm{\sigma}^{\text{QD}_i}=(
\sigma_x^{\text{QD}_i},
\sigma_y^{\text{QD}_i},
\sigma_z^{\text{QD}_i}
)$
are the Pauli operators for the electrons at QD$_i$. 
For the HQ, this term gives the contribution
\begin{align}
\mathcal{H}^{\delta \bm{B}}_{\left\{\ket{\overline{1}},\ket{\overline{0}}\right\}}=
\delta_B\sigma_x,
\label{eq:Hyp}
\end{align}
with $\delta_B=\frac{g\mu_B}{2}\left[
-\frac{3+\cos^2\left(\frac{\theta}{2}\right)}{6}\delta B^{\text{QD}_1}_z
+\frac{8\cos^2\left(\frac{\theta}{2}\right)}{6}\delta B^{\text{QD}_2}_z
\right]$ and $\sigma_x=\op{\overline{1}}{\overline{0}}+\op{\overline{0}}{\overline{1}}$.

The nuclear magnetic field can be treated as static during a single gate operation, but it fluctuates between successive measurements. Typical magnitudes of $\delta B_z^{\text{QD}}$ are $5~\text{mT}$ for GaAs QDs and $100~\mu\text{T}$ for Si QDs. The associate frequencies are $g\mu_B \delta B_z^{\text{QD}}/h\approx 30~\text{MHz}$ for GaAs QDs and $g\mu_B \delta B_z^{\text{QD}}/h\approx 3~\text{MHz}$ for Si QDs, while $\overline{\Delta}_0/h$ reaches several GHz. Therefore, one can treat \eref{eq:Hyp} as a small perturbation to \eref{eq:Ham3}.

\subsection{Charge Noise}
Fluctuating electric fields are caused by the filling and unfilling of charge traps \cite{hu2006,taylor2007} or by fluctuating gate potentials. One can describe the dominant contribution for a DQD qubit by introducing an uncertainty in the detuning parameter $\epsilon$,\cite{dial2013} which causes the noise term
\begin{align}
\mathcal{H}^{\delta \epsilon}_{\left\{\ket{\overline{1}},\ket{\overline{0}}\right\}}=
\frac{\delta^2}{2\left(\delta^2+\Delta_1^2\right)}\delta\epsilon~\sigma_x
\label{eq:CNoise}
\end{align}
for the HQ at the sweet spot $\epsilon^*$, with $\sigma_x=\op{\overline{1}}{\overline{0}}+\op{\overline{0}}{\overline{1}}$.

Charge fluctuation are slow compared to the gate times. The rms values of $\delta \epsilon$ reach a few $\mu \text{eV}$ ($1~\mu eV/h\approx 0.24 ~\text{GHz}$) in GaAs \cite{petersson2010,dial2013,medford2013-2} and Si \cite{shi2013}. The factor $\frac{\delta^2}{2\left(\delta^2+\Delta_1^2\right)}$ in \eref{eq:CNoise} gives an algebraic suppression to charge noise compared to a charge qubit. Note that this suppression is much smaller than for spin qubits \cite{taylor2007}. The parameters of the previous analysis suggest magnitudes of the fluctuations $\frac{\delta^2}{2\left(\delta^2+\Delta_1^2\right)}1~\mu\text{eV}/h\approx 76~\text{MHz}$. This number it still by two orders of magnitudes smaller than $\overline{\Delta}_0/h\approx 4.5~\text{GHz}$.

\section{\label{sec:Summ}
Summary and Conclusion}

This article has analyzed quantum computation for the HQ. The HQ is an exchange-only qubit,\cite{divincenzo2000} where three electrons are confined at a DQD.\cite{shi2012} There are two avoided level crossings between states in $\left(2,1\right)$ and $\left(1,2\right)$ in the transition region of these charge configurations. Controlled transfers through the avoided crossing have realized single-qubit gates experimentally\cite{shi2014,kim2014,kim2014-2}, while the possibility of resonant gates motivates the search for optimal operation points.\cite{kim2014-2,taylor2013,medford2013,mehl2014-2} I introduce such an optimal operation point at a sweet spot that is exceptionally noise insensitive. All the gate operations can be done close to this sweet spot.

This paper has derived an effective qubit description at the sweet spot. Two HQs can be coupled via Coulomb and exchange interactions. If both HQs are operated at their sweet spots, then the Coulomb interaction stays constant but the inter-qubit exchange interaction can be manipulated quickly. In a first approach, two qubits of similar eigenfrequencies are analyzed. The exchange interaction can bring the qubits in and out of resonance. Two qubits of identical eigenfrequencies entangle under static time evolutions, but two distinct qubits evolve independently. In a second approach, two highly distinct qubits are analyzed. The Coulomb interaction stays constant, and only resonant electric signals are needed to realize single-qubit and two-qubit gates.

This paper has simulated HQs in Si and GaAs with realistic parameters that are extracted from experiments.\cite{shi2014,kim2014,kim2014-2} Most critically, the gate operations require Coulomb interactions between HQs that are small compared to the naive estimates for the setup. Especially, if inter-qubit exchange interactions are used for the entangling operations, then the inter-qubit Coulomb couplings should be small. This paper has discussed how weak inter-qubit Coulomb couplings can be realized. It should be possible to realize such weak inter-qubit couplings with a careful design of the QD layout.

Fault-tolerant quantum computation requires high-fidelity quantum gates with error probabilities below $1\%$ \cite{fowler2012,jones2012}. My simulations showed that quantum computation with this infidelities is possible if multi-qubit arrangements of HQ with the described parameters can be fabricated. Nuclear spin noise and charge noise are less critical. Additionally, these statistical errors can be reduced with refocusing protocols with similar approaches as for DQDs\cite{barthel2010-1,bluhm2011,medford2012} and TQDs\cite{medford2013,medford2013-2}. A reduction of the nuclear spin fluctuation can be realized by preparing the nuclear spin bath\cite{foletti2009,bluhm2010} or with QD materials that contain nuclei of zero spins\cite{veldhorst2014,veldhorst2014-2} if the need arises.

A recent experiment also realized resonant single-qubit gates for the HQ.\cite{kim2015} In this case, the HQ is operated deep in $\left(1,2\right)$ which reduces the influence of charge noise. Resonant single-qubit gates are possible when the transition region to $\left(2,1\right)$ is approached, while one still stays away from the anticrossings. Two-qubit gates with exchange interactions cannot be used to entangle HQs in \rcite{kim2015} in the same way as in the study of this paper because many leakage states are degenerate with the computational states. In \rcite{kim2015}, the $S=3/2$, $s_z=1/2$ state is nearly degenerate with the qubit states (cf. \fref{fig:03}). The Coulomb interaction can still be used to entangle HQs, similar to \sref{sec:Cap}, while the interaction Hamiltonian is $\propto\sigma_z^{\text{L}}\sigma_z^{\text{R}}$. Universal qubit control likely requires DC control of the interaction Hamiltonian. Either two HQs are isolated from each other, or they are coupled via $\sigma_z^{\text{L}}\sigma_z^{\text{R}}$ (in contrast to the approach in \sref{sec:UnivB} where single-qubit and two-qubit gates are possible without changing the operation points of the HQs).

Overall, the HQ is an interesting candidate for further experimental and theoretical studies. It has many characteristics of a charge qubit, especially with its fast operation times. On the other hand, it can be protected from charge noise similar to a spin qubit. The described sweet spot manipulations classify the HQ as a mixture of a charge qubit and a spin qubit that has advantages from both setups. This paper has shown that universal gate operations can be realized for the HQ at the threshold of quantum error correction, which should further motivate the search for optimal manipulation protocols of HQs.

\begin{acknowledgements}
I am grateful for support from the Alexander von Humboldt foundation.
\end{acknowledgements}

\begin{appendix}

\section{\label{app:Fid}
Gate Fidelities}

The fidelity that is used in the paper should be defined in the following. To characterize gate fidelities, the time evolution $U$ is compared to the ideal time evolution $U_{\mathcal{I}}$. The state space is doubled to two identical Hilbert spaces $\text{R}$ and $\text{S}$. The entanglement fidelity \cite{nielsen2000}
\begin{align}
F=\text{tr}
\left\{\rho^{\text{RS}}
\bm{1}^{\text{R}}\otimes\left[U_{\mathcal{I}}^{-1}U\right]^{\text{S}}
\rho^{\text{RS}}
\bm{1}^{\text{R}}\otimes\left[U^{-1}U_{\mathcal{I}}\right]^{\text{S}}
\right\}
\label{eq:Fidelity}
\end{align}
is a measure for the gate performance. \eref{eq:Fidelity} compares the time evolution $\left[U_{\mathcal{I}}^{-1}U\right]^{\text{S}}$ of $\text{S}$ with a reference system $\text{R}$ that is unchanged. $\rho^{\text{RS}}=\op{\psi}{\psi}$ is a maximally entangled state of the combined Hilbert space, e.g., $\ket{\psi}=\left(\ket{0000}+\ket{0110}+\ket{1001}+\ket{1111}\right)/2$.

\section{\label{app:Exc}
Effective Inter-Qubit Exchange Hamiltonian}

This section describes the derivation of the effective interactions from \sref{sec:Exc} for weak tunnel couplings between HQ\tu{L} and HQ\tu{R}. The HQs are operated at their sweet spots. The dominant Hamiltonian $\mathcal{H}$ is determined by \eref{eq:Ham1} [with the approximations in \eref{eq:Ham3}]. \eref{eq:Pert} defines the coupling between QD$_{2}^{\text{L}}$ and QD$_{1}^{\text{R}}$ (called $\mathcal{H}^\prime$). The system is tuned towards $\left(1,1,2,2\right)$, while the computational states still remain in the ground states configurations. The low-energy subspace $\mathcal{P}=\mathcal{P}_{\mathcal{Q}}+\mathcal{P}_{\mathcal{L}}$ contains the qubit states $\mathcal{P}_{\mathcal{Q}}=\left\{\ket{\overline{1}^{\text{L}}\overline{1}^{\text{R}}},
\ket{\overline{1}^{\text{L}}\overline{0}^{\text{R}}},
\ket{\overline{0}^{\text{L}}\overline{1}^{\text{R}}},
\ket{\overline{0}^{\text{L}}\overline{0}^{\text{R}}}\right\}$. The leakage subspace contains the states $\mathcal{P}_{\mathcal{L}}=\left\{\ket{\alpha},\ket{\beta},\ket{\gamma},\ket{\delta},\ket{\nu}\right\}$ that are defined in \tref{tab:01}.

Only states that have a singlet at QD$_1^{\text{R}}$ are included in $\left(1,1,2,2\right)$:
$\mathcal{Q}=\{\ket{\uparrow\uparrow \text{S S}}, \ket{\uparrow\uparrow \text{S T}_0}, \ket{\uparrow\downarrow \text{S T}_+}, 
\ket{\downarrow\uparrow \text{S T}_+}\}$. A weak tuning towards $\left(1,1,2,2\right)$ only occupies $\mathcal{Q}$ virtually.
\eref{eq:Pert} couples states in $\left(1,2,1,2\right)$ and $\left(1,1,2,2\right)$. Schrieffer-Wolff perturbation theory constructs an effective Hamiltonian on $\mathcal{P}$, assuming that the coupling between $\mathcal{P}$ and $\mathcal{Q}$ is weak. Additionally, $\mathcal{Q}$ has higher energy than $\mathcal{P}$.\cite{schrieffer1966,winkler2010,bravyi2011}

The effective interaction on $\mathcal{P}$ is defined in second-order Schrieffer-Wolff perturbation theory by
\begin{align}
\label{eq:SO-SW}
\mathcal{H}_{\mathcal{P}}^{ij}&=\mathcal{H}_\mathcal{P}^{ij}
\\\nonumber&+
\frac{1}{2}\sum_{k\in\mathcal{Q}}
\left(\mathcal{H}^\prime_{\mathcal{P}\mathcal{Q}}\right)^{ik}
\left(\frac{1}{E_i-E_k}+\frac{1}{E_j-E_k}\right)
\left(\mathcal{H}^\prime_{\mathcal{Q}\mathcal{P}}\right)^{ik}.
\end{align}
\eref{eq:SO-SW} uses the transition matrix element between the states $\ket{i}$ and $\ket{j}$ from $\mathcal{P}$: $\mathcal{H}_{\mathcal{P}}^{ij}=\Dirac{i}{\mathcal{H}}{j}$.
$\left(\mathcal{H}^\prime_{\mathcal{P}\mathcal{Q}}\right)^{ik}$ and
$\left(\mathcal{H}^\prime_{\mathcal{Q}\mathcal{P}}\right)^{ki}$ are the transition matrix elements between the states $\ket{i}$ from $\mathcal{P}$ and $\ket{k}$ from $\mathcal{Q}$. $E_i$ is the energy of the state $\ket{i}$.

In the whole paper, it is assumed that the energy difference between $\mathcal{P}$ and $\mathcal{Q}$ is large, especially much larger than the energy differences between states in $\mathcal{P}$ or between states in $\mathcal{Q}$. One can therefore use for all $\ket{i}\in\mathcal{P}$ and $\ket{k}\in\mathcal{Q}$: $E_i-E_k\approx
E_{\left(1,2,1,2\right)}-E_{\left(1,1,2,2\right)}$.

\end{appendix}

\bibliography{library}

\begin{thebibliography}{80}%
\makeatletter
\providecommand \@ifxundefined [1]{%
 \@ifx{#1\undefined}
}%
\providecommand \@ifnum [1]{%
 \ifnum #1\expandafter \@firstoftwo
 \else \expandafter \@secondoftwo
 \fi
}%
\providecommand \@ifx [1]{%
 \ifx #1\expandafter \@firstoftwo
 \else \expandafter \@secondoftwo
 \fi
}%
\providecommand \natexlab [1]{#1}%
\providecommand \enquote  [1]{``#1''}%
\providecommand \bibnamefont  [1]{#1}%
\providecommand \bibfnamefont [1]{#1}%
\providecommand \citenamefont [1]{#1}%
\providecommand \href@noop [0]{\@secondoftwo}%
\providecommand \href [0]{\begingroup \@sanitize@url \@href}%
\providecommand \@href[1]{\@@startlink{#1}\@@href}%
\providecommand \@@href[1]{\endgroup#1\@@endlink}%
\providecommand \@sanitize@url [0]{\catcode `\\12\catcode `\$12\catcode
  `\&12\catcode `\#12\catcode `\^12\catcode `\_12\catcode `\%12\relax}%
\providecommand \@@startlink[1]{}%
\providecommand \@@endlink[0]{}%
\providecommand \url  [0]{\begingroup\@sanitize@url \@url }%
\providecommand \@url [1]{\endgroup\@href {#1}{\urlprefix }}%
\providecommand \urlprefix  [0]{URL }%
\providecommand \Eprint [0]{\href }%
\providecommand \doibase [0]{http://dx.doi.org/}%
\providecommand \selectlanguage [0]{\@gobble}%
\providecommand \bibinfo  [0]{\@secondoftwo}%
\providecommand \bibfield  [0]{\@secondoftwo}%
\providecommand \translation [1]{[#1]}%
\providecommand \BibitemOpen [0]{}%
\providecommand \bibitemStop [0]{}%
\providecommand \bibitemNoStop [0]{.\EOS\space}%
\providecommand \EOS [0]{\spacefactor3000\relax}%
\providecommand \BibitemShut  [1]{\csname bibitem#1\endcsname}%
\let\auto@bib@innerbib\@empty
\bibitem [{\citenamefont {Loss}\ and\ \citenamefont
  {DiVincenzo}(1998)}]{loss1998}%
  \BibitemOpen
  \bibfield  {author} {\bibinfo {author} {\bibfnamefont {D.}~\bibnamefont
  {Loss}}\ and\ \bibinfo {author} {\bibfnamefont {D.~P.}\ \bibnamefont
  {DiVincenzo}},\ }\bibfield  {title} {\enquote {\bibinfo {title} {Quantum
  computation with quantum dots},}\ }\href {\doibase 10.1103/PhysRevA.57.120}
  {\bibfield  {journal} {\bibinfo  {journal} {Phys. Rev. A}\ }\textbf {\bibinfo
  {volume} {57}},\ \bibinfo {pages} {120} (\bibinfo {year} {1998})}\BibitemShut
  {NoStop}%
\bibitem [{\citenamefont {Kane}(1998)}]{kane1998}%
  \BibitemOpen
  \bibfield  {author} {\bibinfo {author} {\bibfnamefont {B.~E.}\ \bibnamefont
  {Kane}},\ }\bibfield  {title} {\enquote {\bibinfo {title} {A silicon-based
  nuclear spin quantum computer},}\ }\href {\doibase 10.1038/30156} {\bibfield
  {journal} {\bibinfo  {journal} {Nature (London)}\ }\textbf {\bibinfo {volume}
  {393}},\ \bibinfo {pages} {133} (\bibinfo {year} {1998})}\BibitemShut
  {NoStop}%
\bibitem [{\citenamefont {Awschalom}(2002)}]{awschalom2002}%
  \BibitemOpen
  \bibfield  {author} {\bibinfo {author} {\bibfnamefont {D.~D.}\ \bibnamefont
  {Awschalom}},\ }\href@noop {} {\emph {\bibinfo {title} {{Semiconductor
  Spintronics and Quantum Computation}}}},\ Nanoscience and technology\
  (\bibinfo  {publisher} {Springer},\ \bibinfo {address} {Berlin},\ \bibinfo
  {year} {2002})\BibitemShut {NoStop}%
\bibitem [{\citenamefont {Levy}(2002)}]{levy2002}%
  \BibitemOpen
  \bibfield  {author} {\bibinfo {author} {\bibfnamefont {J.}~\bibnamefont
  {Levy}},\ }\bibfield  {title} {\enquote {\bibinfo {title} {{Universal Quantum
  Computation with Spin-1/2 Pairs and Heisenberg Exchange}},}\ }\href {\doibase
  10.1103/PhysRevLett.89.147902} {\bibfield  {journal} {\bibinfo  {journal}
  {Phys. Rev. Lett.}\ }\textbf {\bibinfo {volume} {89}},\ \bibinfo {pages}
  {147902} (\bibinfo {year} {2002})}\BibitemShut {NoStop}%
\bibitem [{\citenamefont {Taylor}\ \emph {et~al.}(2005)\citenamefont {Taylor},
  \citenamefont {Engel}, \citenamefont {D{\"u}r}, \citenamefont {Yacoby},
  \citenamefont {Marcus}, \citenamefont {Zoller},\ and\ \citenamefont
  {Lukin}}]{taylor2005}%
  \BibitemOpen
  \bibfield  {author} {\bibinfo {author} {\bibfnamefont {J.~M.}\ \bibnamefont
  {Taylor}}, \bibinfo {author} {\bibfnamefont {H.-A.}\ \bibnamefont {Engel}},
  \bibinfo {author} {\bibfnamefont {W.}~\bibnamefont {D{\"u}r}}, \bibinfo
  {author} {\bibfnamefont {A.}~\bibnamefont {Yacoby}}, \bibinfo {author}
  {\bibfnamefont {C.~M.}\ \bibnamefont {Marcus}}, \bibinfo {author}
  {\bibfnamefont {P.}~\bibnamefont {Zoller}}, \ and\ \bibinfo {author}
  {\bibfnamefont {M.~D.}\ \bibnamefont {Lukin}},\ }\bibfield  {title} {\enquote
  {\bibinfo {title} {Fault-tolerant architecture for quantum computation using
  electrically controlled semiconductor spins},}\ }\href {\doibase
  10.1038/nphys174} {\bibfield  {journal} {\bibinfo  {journal} {Nat. Phys.}\
  }\textbf {\bibinfo {volume} {1}},\ \bibinfo {pages} {177} (\bibinfo {year}
  {2005})}\BibitemShut {NoStop}%
\bibitem [{\citenamefont {DiVincenzo}\ \emph {et~al.}(2000)\citenamefont
  {DiVincenzo}, \citenamefont {Bacon}, \citenamefont {Kempe}, \citenamefont
  {Burkard},\ and\ \citenamefont {Whaley}}]{divincenzo2000}%
  \BibitemOpen
  \bibfield  {author} {\bibinfo {author} {\bibfnamefont {D.~P.}\ \bibnamefont
  {DiVincenzo}}, \bibinfo {author} {\bibfnamefont {D.}~\bibnamefont {Bacon}},
  \bibinfo {author} {\bibfnamefont {J.}~\bibnamefont {Kempe}}, \bibinfo
  {author} {\bibfnamefont {G.}~\bibnamefont {Burkard}}, \ and\ \bibinfo
  {author} {\bibfnamefont {K.~B.}\ \bibnamefont {Whaley}},\ }\bibfield  {title}
  {\enquote {\bibinfo {title} {Universal quantum computation with the exchange
  interaction},}\ }\href {\doibase 10.1038/35042541} {\bibfield  {journal}
  {\bibinfo  {journal} {Nature (London)}\ }\textbf {\bibinfo {volume} {408}},\
  \bibinfo {pages} {339} (\bibinfo {year} {2000})}\BibitemShut {NoStop}%
\bibitem [{\citenamefont {Bacon}\ \emph {et~al.}(2000)\citenamefont {Bacon},
  \citenamefont {Kempe}, \citenamefont {Lidar},\ and\ \citenamefont
  {Whaley}}]{bacon2000}%
  \BibitemOpen
  \bibfield  {author} {\bibinfo {author} {\bibfnamefont {D.}~\bibnamefont
  {Bacon}}, \bibinfo {author} {\bibfnamefont {J.}~\bibnamefont {Kempe}},
  \bibinfo {author} {\bibfnamefont {D.~A.}\ \bibnamefont {Lidar}}, \ and\
  \bibinfo {author} {\bibfnamefont {K.~B.}\ \bibnamefont {Whaley}},\ }\bibfield
   {title} {\enquote {\bibinfo {title} {{Universal Fault-Tolerant Quantum
  Computation on Decoherence-Free Subspaces}},}\ }\href {\doibase
  10.1103/PhysRevLett.85.1758} {\bibfield  {journal} {\bibinfo  {journal}
  {Phys. Rev. Lett.}\ }\textbf {\bibinfo {volume} {85}},\ \bibinfo {pages}
  {1758} (\bibinfo {year} {2000})}\BibitemShut {NoStop}%
\bibitem [{\citenamefont {Kempe}\ \emph {et~al.}(2001)\citenamefont {Kempe},
  \citenamefont {Bacon}, \citenamefont {Lidar},\ and\ \citenamefont
  {Whaley}}]{kempe2001}%
  \BibitemOpen
  \bibfield  {author} {\bibinfo {author} {\bibfnamefont {J.}~\bibnamefont
  {Kempe}}, \bibinfo {author} {\bibfnamefont {D.}~\bibnamefont {Bacon}},
  \bibinfo {author} {\bibfnamefont {D.~A.}\ \bibnamefont {Lidar}}, \ and\
  \bibinfo {author} {\bibfnamefont {K.~B.}\ \bibnamefont {Whaley}},\ }\bibfield
   {title} {\enquote {\bibinfo {title} {Theory of decoherence-free
  fault-tolerant universal quantum computation},}\ }\href {\doibase
  10.1103/PhysRevA.63.042307} {\bibfield  {journal} {\bibinfo  {journal} {Phys.
  Rev. A}\ }\textbf {\bibinfo {volume} {63}},\ \bibinfo {pages} {042307}
  (\bibinfo {year} {2001})}\BibitemShut {NoStop}%
\bibitem [{\citenamefont {Shi}\ \emph {et~al.}(2012)\citenamefont {Shi},
  \citenamefont {Simmons}, \citenamefont {Prance}, \citenamefont {Gamble},
  \citenamefont {Koh}, \citenamefont {Shim}, \citenamefont {Hu}, \citenamefont
  {Savage}, \citenamefont {Lagally}, \citenamefont {Eriksson}, \citenamefont
  {Friesen},\ and\ \citenamefont {Coppersmith}}]{shi2012}%
  \BibitemOpen
  \bibfield  {author} {\bibinfo {author} {\bibfnamefont {Z.}~\bibnamefont
  {Shi}}, \bibinfo {author} {\bibfnamefont {C.~B.}\ \bibnamefont {Simmons}},
  \bibinfo {author} {\bibfnamefont {J.~R.}\ \bibnamefont {Prance}}, \bibinfo
  {author} {\bibfnamefont {J.~K.}\ \bibnamefont {Gamble}}, \bibinfo {author}
  {\bibfnamefont {T.~S.}\ \bibnamefont {Koh}}, \bibinfo {author} {\bibfnamefont
  {Y.-P.}\ \bibnamefont {Shim}}, \bibinfo {author} {\bibfnamefont
  {X.}~\bibnamefont {Hu}}, \bibinfo {author} {\bibfnamefont {D.~E.}\
  \bibnamefont {Savage}}, \bibinfo {author} {\bibfnamefont {M.~G.}\
  \bibnamefont {Lagally}}, \bibinfo {author} {\bibfnamefont {M.~A.}\
  \bibnamefont {Eriksson}}, \bibinfo {author} {\bibfnamefont {M.}~\bibnamefont
  {Friesen}}, \ and\ \bibinfo {author} {\bibfnamefont {S.~N.}\ \bibnamefont
  {Coppersmith}},\ }\bibfield  {title} {\enquote {\bibinfo {title} {{Fast
  Hybrid Silicon Double-Quantum-Dot Qubit}},}\ }\href {\doibase
  10.1103/PhysRevLett.108.140503} {\bibfield  {journal} {\bibinfo  {journal}
  {Phys. Rev. Lett.}\ }\textbf {\bibinfo {volume} {108}},\ \bibinfo {pages}
  {140503} (\bibinfo {year} {2012})}\BibitemShut {NoStop}%
\bibitem [{\citenamefont {Petta}\ \emph {et~al.}(2005)\citenamefont {Petta},
  \citenamefont {Johnson}, \citenamefont {Taylor}, \citenamefont {Laird},
  \citenamefont {Yacoby}, \citenamefont {Lukin}, \citenamefont {Marcus},
  \citenamefont {Hanson},\ and\ \citenamefont {Gossard}}]{petta2005}%
  \BibitemOpen
  \bibfield  {author} {\bibinfo {author} {\bibfnamefont {J.~R.}\ \bibnamefont
  {Petta}}, \bibinfo {author} {\bibfnamefont {A.~C.}\ \bibnamefont {Johnson}},
  \bibinfo {author} {\bibfnamefont {J.~M.}\ \bibnamefont {Taylor}}, \bibinfo
  {author} {\bibfnamefont {E.~A.}\ \bibnamefont {Laird}}, \bibinfo {author}
  {\bibfnamefont {A.}~\bibnamefont {Yacoby}}, \bibinfo {author} {\bibfnamefont
  {M.~D.}\ \bibnamefont {Lukin}}, \bibinfo {author} {\bibfnamefont {C.~M.}\
  \bibnamefont {Marcus}}, \bibinfo {author} {\bibfnamefont {M.~P.}\
  \bibnamefont {Hanson}}, \ and\ \bibinfo {author} {\bibfnamefont {A.~C.}\
  \bibnamefont {Gossard}},\ }\bibfield  {title} {\enquote {\bibinfo {title}
  {{Coherent Manipulation of Coupled Electron Spins in Semiconductor Quantum
  Dots}},}\ }\href {\doibase 10.1126/science.1116955} {\bibfield  {journal}
  {\bibinfo  {journal} {Science}\ }\textbf {\bibinfo {volume} {309}},\ \bibinfo
  {pages} {2180} (\bibinfo {year} {2005})}\BibitemShut {NoStop}%
\bibitem [{\citenamefont {Maune}\ \emph {et~al.}(2012)\citenamefont {Maune},
  \citenamefont {Borselli}, \citenamefont {Huang}, \citenamefont {Ladd},
  \citenamefont {Deelman}, \citenamefont {Holabird}, \citenamefont {Kiselev},
  \citenamefont {Alvarado-Rodriguez}, \citenamefont {Ross}, \citenamefont
  {Schmitz}, \citenamefont {Sokolich}, \citenamefont {Watson}, \citenamefont
  {Gyure},\ and\ \citenamefont {Hunter}}]{maune2012}%
  \BibitemOpen
  \bibfield  {author} {\bibinfo {author} {\bibfnamefont {B.~M.}\ \bibnamefont
  {Maune}}, \bibinfo {author} {\bibfnamefont {M.~G.}\ \bibnamefont {Borselli}},
  \bibinfo {author} {\bibfnamefont {B.}~\bibnamefont {Huang}}, \bibinfo
  {author} {\bibfnamefont {T.~D.}\ \bibnamefont {Ladd}}, \bibinfo {author}
  {\bibfnamefont {P.~W.}\ \bibnamefont {Deelman}}, \bibinfo {author}
  {\bibfnamefont {K.~S.}\ \bibnamefont {Holabird}}, \bibinfo {author}
  {\bibfnamefont {A.~A.}\ \bibnamefont {Kiselev}}, \bibinfo {author}
  {\bibfnamefont {I.}~\bibnamefont {Alvarado-Rodriguez}}, \bibinfo {author}
  {\bibfnamefont {R.~S.}\ \bibnamefont {Ross}}, \bibinfo {author}
  {\bibfnamefont {A.~E.}\ \bibnamefont {Schmitz}}, \bibinfo {author}
  {\bibfnamefont {M.}~\bibnamefont {Sokolich}}, \bibinfo {author}
  {\bibfnamefont {C.~A.}\ \bibnamefont {Watson}}, \bibinfo {author}
  {\bibfnamefont {M.~F.}\ \bibnamefont {Gyure}}, \ and\ \bibinfo {author}
  {\bibfnamefont {A.~T.}\ \bibnamefont {Hunter}},\ }\bibfield  {title}
  {\enquote {\bibinfo {title} {Coherent singlet-triplet oscillations in a
  silicon-based double quantum dot},}\ }\href {\doibase 10.1038/nature10707}
  {\bibfield  {journal} {\bibinfo  {journal} {Nature (London)}\ }\textbf
  {\bibinfo {volume} {481}},\ \bibinfo {pages} {344} (\bibinfo {year}
  {2012})}\BibitemShut {NoStop}%
\bibitem [{\citenamefont {Koppens}\ \emph {et~al.}(2006)\citenamefont
  {Koppens}, \citenamefont {Buizert}, \citenamefont {Tielrooij}, \citenamefont
  {Vink}, \citenamefont {Nowack}, \citenamefont {Meunier}, \citenamefont
  {Kouwenhoven},\ and\ \citenamefont {Vandersypen}}]{koppens2006}%
  \BibitemOpen
  \bibfield  {author} {\bibinfo {author} {\bibfnamefont {F.~H.~L.}\
  \bibnamefont {Koppens}}, \bibinfo {author} {\bibfnamefont {C.}~\bibnamefont
  {Buizert}}, \bibinfo {author} {\bibfnamefont {K.-J.}\ \bibnamefont
  {Tielrooij}}, \bibinfo {author} {\bibfnamefont {I.~T.}\ \bibnamefont {Vink}},
  \bibinfo {author} {\bibfnamefont {K.~C.}\ \bibnamefont {Nowack}}, \bibinfo
  {author} {\bibfnamefont {T.}~\bibnamefont {Meunier}}, \bibinfo {author}
  {\bibfnamefont {L.~P.}\ \bibnamefont {Kouwenhoven}}, \ and\ \bibinfo {author}
  {\bibfnamefont {L.~M.~K.}\ \bibnamefont {Vandersypen}},\ }\bibfield  {title}
  {\enquote {\bibinfo {title} {Driven coherent oscillations of a single
  electron spin in a quantum dot},}\ }\href {\doibase 10.1038/nature05065}
  {\bibfield  {journal} {\bibinfo  {journal} {Nature (London)}\ }\textbf
  {\bibinfo {volume} {442}},\ \bibinfo {pages} {766} (\bibinfo {year}
  {2006})}\BibitemShut {NoStop}%
\bibitem [{\citenamefont {Pla}\ \emph {et~al.}(2012)\citenamefont {Pla},
  \citenamefont {Tan}, \citenamefont {Dehollain}, \citenamefont {Lim},
  \citenamefont {Morton}, \citenamefont {Jamieson}, \citenamefont {Dzurak},\
  and\ \citenamefont {Morello}}]{pla2012}%
  \BibitemOpen
  \bibfield  {author} {\bibinfo {author} {\bibfnamefont {J.~J.}\ \bibnamefont
  {Pla}}, \bibinfo {author} {\bibfnamefont {K.~Y.}\ \bibnamefont {Tan}},
  \bibinfo {author} {\bibfnamefont {J.~P.}\ \bibnamefont {Dehollain}}, \bibinfo
  {author} {\bibfnamefont {W.~H.}\ \bibnamefont {Lim}}, \bibinfo {author}
  {\bibfnamefont {J.~J.~L.}\ \bibnamefont {Morton}}, \bibinfo {author}
  {\bibfnamefont {D.~N.}\ \bibnamefont {Jamieson}}, \bibinfo {author}
  {\bibfnamefont {A.~S.}\ \bibnamefont {Dzurak}}, \ and\ \bibinfo {author}
  {\bibfnamefont {A.}~\bibnamefont {Morello}},\ }\bibfield  {title} {\enquote
  {\bibinfo {title} {A single-atom electron spin qubit in silicon},}\ }\href
  {\doibase 10.1038/nature11449} {\bibfield  {journal} {\bibinfo  {journal}
  {Nature (London)}\ }\textbf {\bibinfo {volume} {489}},\ \bibinfo {pages}
  {541} (\bibinfo {year} {2012})}\BibitemShut {NoStop}%
\bibitem [{\citenamefont {Veldhorst}\ \emph
  {et~al.}(2014{\natexlab{a}})\citenamefont {Veldhorst}, \citenamefont {Hwang},
  \citenamefont {Yang}, \citenamefont {Leenstra}, \citenamefont {{de Ronde}},
  \citenamefont {Dehollain}, \citenamefont {Muhonen}, \citenamefont {Hudson},
  \citenamefont {Itoh}, \citenamefont {Morello},\ and\ \citenamefont
  {Dzurak}}]{veldhorst2014}%
  \BibitemOpen
  \bibfield  {author} {\bibinfo {author} {\bibfnamefont {M.}~\bibnamefont
  {Veldhorst}}, \bibinfo {author} {\bibfnamefont {J.~C.~C.}\ \bibnamefont
  {Hwang}}, \bibinfo {author} {\bibfnamefont {C.~H.}\ \bibnamefont {Yang}},
  \bibinfo {author} {\bibfnamefont {A.~W.}\ \bibnamefont {Leenstra}}, \bibinfo
  {author} {\bibfnamefont {B.}~\bibnamefont {{de Ronde}}}, \bibinfo {author}
  {\bibfnamefont {J.~P.}\ \bibnamefont {Dehollain}}, \bibinfo {author}
  {\bibfnamefont {J.~T.}\ \bibnamefont {Muhonen}}, \bibinfo {author}
  {\bibfnamefont {F.~E.}\ \bibnamefont {Hudson}}, \bibinfo {author}
  {\bibfnamefont {K.~M.}\ \bibnamefont {Itoh}}, \bibinfo {author}
  {\bibfnamefont {A.}~\bibnamefont {Morello}}, \ and\ \bibinfo {author}
  {\bibfnamefont {A.~S.}\ \bibnamefont {Dzurak}},\ }\bibfield  {title}
  {\enquote {\bibinfo {title} {An addressable quantum dot qubit with
  fault-tolerant control-fidelity},}\ }\href {\doibase 10.1038/nnano.2014.216}
  {\bibfield  {journal} {\bibinfo  {journal} {Nat. Nanotechnol.}\ }\textbf
  {\bibinfo {volume} {9}},\ \bibinfo {pages} {981} (\bibinfo {year}
  {2014}{\natexlab{a}})}\BibitemShut {NoStop}%
\bibitem [{\citenamefont {Nowack}\ \emph {et~al.}(2007)\citenamefont {Nowack},
  \citenamefont {Koppens}, \citenamefont {Nazarov},\ and\ \citenamefont
  {Vandersypen}}]{nowack2007}%
  \BibitemOpen
  \bibfield  {author} {\bibinfo {author} {\bibfnamefont {K.~C.}\ \bibnamefont
  {Nowack}}, \bibinfo {author} {\bibfnamefont {F.~H.~L.}\ \bibnamefont
  {Koppens}}, \bibinfo {author} {\bibfnamefont {Y.~V.}\ \bibnamefont
  {Nazarov}}, \ and\ \bibinfo {author} {\bibfnamefont {L.~M.~K.}\ \bibnamefont
  {Vandersypen}},\ }\bibfield  {title} {\enquote {\bibinfo {title} {{Coherent
  Control of a Single Electron Spin with Electric Fields}},}\ }\href {\doibase
  10.1126/science.1148092} {\bibfield  {journal} {\bibinfo  {journal}
  {Science}\ }\textbf {\bibinfo {volume} {318}},\ \bibinfo {pages} {1430}
  (\bibinfo {year} {2007})}\BibitemShut {NoStop}%
\bibitem [{\citenamefont {Kawakami}\ \emph {et~al.}(2014)\citenamefont
  {Kawakami}, \citenamefont {Scarlino}, \citenamefont {Ward}, \citenamefont
  {Braakman}, \citenamefont {Savage}, \citenamefont {Lagally}, \citenamefont
  {Friesen}, \citenamefont {Coppersmith}, \citenamefont {Eriksson},\ and\
  \citenamefont {Vandersypen}}]{kawakami2014}%
  \BibitemOpen
  \bibfield  {author} {\bibinfo {author} {\bibfnamefont {E.}~\bibnamefont
  {Kawakami}}, \bibinfo {author} {\bibfnamefont {P.}~\bibnamefont {Scarlino}},
  \bibinfo {author} {\bibfnamefont {D.~R.}\ \bibnamefont {Ward}}, \bibinfo
  {author} {\bibfnamefont {F.~R.}\ \bibnamefont {Braakman}}, \bibinfo {author}
  {\bibfnamefont {D.~E.}\ \bibnamefont {Savage}}, \bibinfo {author}
  {\bibfnamefont {M.~G.}\ \bibnamefont {Lagally}}, \bibinfo {author}
  {\bibfnamefont {M.}~\bibnamefont {Friesen}}, \bibinfo {author} {\bibfnamefont
  {S.~N.}\ \bibnamefont {Coppersmith}}, \bibinfo {author} {\bibfnamefont
  {M.~A.}\ \bibnamefont {Eriksson}}, \ and\ \bibinfo {author} {\bibfnamefont
  {L.~M.~K.}\ \bibnamefont {Vandersypen}},\ }\bibfield  {title} {\enquote
  {\bibinfo {title} {{Electrical control of a long-lived spin qubit in a
  Si/SiGe quantum dot}},}\ }\href {\doibase 10.1038/nnano.2014.153} {\bibfield
  {journal} {\bibinfo  {journal} {Nat. Nanotechnol.}\ }\textbf {\bibinfo
  {volume} {9}},\ \bibinfo {pages} {666} (\bibinfo {year} {2014})}\BibitemShut
  {NoStop}%
\bibitem [{\citenamefont {Yoneda}\ \emph {et~al.}(2014)\citenamefont {Yoneda},
  \citenamefont {Otsuka}, \citenamefont {Nakajima}, \citenamefont {Takakura},
  \citenamefont {Obata}, \citenamefont {Pioro-Ladri{\`e}re}, \citenamefont
  {Lu}, \citenamefont {Palmstr{\o}m}, \citenamefont {Gossard},\ and\
  \citenamefont {Tarucha}}]{yoneda2014}%
  \BibitemOpen
  \bibfield  {author} {\bibinfo {author} {\bibfnamefont {J.}~\bibnamefont
  {Yoneda}}, \bibinfo {author} {\bibfnamefont {T.}~\bibnamefont {Otsuka}},
  \bibinfo {author} {\bibfnamefont {T.}~\bibnamefont {Nakajima}}, \bibinfo
  {author} {\bibfnamefont {T.}~\bibnamefont {Takakura}}, \bibinfo {author}
  {\bibfnamefont {T.}~\bibnamefont {Obata}}, \bibinfo {author} {\bibfnamefont
  {M.}~\bibnamefont {Pioro-Ladri{\`e}re}}, \bibinfo {author} {\bibfnamefont
  {H.}~\bibnamefont {Lu}}, \bibinfo {author} {\bibfnamefont {C.}~\bibnamefont
  {Palmstr{\o}m}}, \bibinfo {author} {\bibfnamefont {A.~C.}\ \bibnamefont
  {Gossard}}, \ and\ \bibinfo {author} {\bibfnamefont {S.}~\bibnamefont
  {Tarucha}},\ }\bibfield  {title} {\enquote {\bibinfo {title} {{Fast
  Electrical Control of Single Electron Spins in Quantum Dots with Vanishing
  Influence from Nuclear Spins}},}\ }\href {\doibase
  10.1103/PhysRevLett.113.267601} {\bibfield  {journal} {\bibinfo  {journal}
  {Phys. Rev. Lett.}\ }\textbf {\bibinfo {volume} {113}},\ \bibinfo {pages}
  {267601} (\bibinfo {year} {2014})}\BibitemShut {NoStop}%
\bibitem [{\citenamefont {Shi}\ \emph {et~al.}(2013)\citenamefont {Shi},
  \citenamefont {Simmons}, \citenamefont {Ward}, \citenamefont {Prance},
  \citenamefont {Mohr}, \citenamefont {Koh}, \citenamefont {Gamble},
  \citenamefont {Wu}, \citenamefont {Savage}, \citenamefont {Lagally},
  \citenamefont {Friesen}, \citenamefont {Coppersmith},\ and\ \citenamefont
  {Eriksson}}]{shi2013}%
  \BibitemOpen
  \bibfield  {author} {\bibinfo {author} {\bibfnamefont {Z.}~\bibnamefont
  {Shi}}, \bibinfo {author} {\bibfnamefont {C.~B.}\ \bibnamefont {Simmons}},
  \bibinfo {author} {\bibfnamefont {D.~R.}\ \bibnamefont {Ward}}, \bibinfo
  {author} {\bibfnamefont {J.~R.}\ \bibnamefont {Prance}}, \bibinfo {author}
  {\bibfnamefont {R.~T.}\ \bibnamefont {Mohr}}, \bibinfo {author}
  {\bibfnamefont {T.~S.}\ \bibnamefont {Koh}}, \bibinfo {author} {\bibfnamefont
  {J.~K.}\ \bibnamefont {Gamble}}, \bibinfo {author} {\bibfnamefont
  {X.}~\bibnamefont {Wu}}, \bibinfo {author} {\bibfnamefont {D.~E.}\
  \bibnamefont {Savage}}, \bibinfo {author} {\bibfnamefont {M.~G.}\
  \bibnamefont {Lagally}}, \bibinfo {author} {\bibfnamefont {M.}~\bibnamefont
  {Friesen}}, \bibinfo {author} {\bibfnamefont {S.~N.}\ \bibnamefont
  {Coppersmith}}, \ and\ \bibinfo {author} {\bibfnamefont {M.~A.}\ \bibnamefont
  {Eriksson}},\ }\bibfield  {title} {\enquote {\bibinfo {title} {{Coherent
  quantum oscillations and echo measurements of a Si charge qubit}},}\ }\href
  {\doibase 10.1103/PhysRevB.88.075416} {\bibfield  {journal} {\bibinfo
  {journal} {Phys. Rev. B}\ }\textbf {\bibinfo {volume} {88}},\ \bibinfo
  {pages} {075416} (\bibinfo {year} {2013})}\BibitemShut {NoStop}%
\bibitem [{\citenamefont {Shi}\ \emph {et~al.}(2014)\citenamefont {Shi},
  \citenamefont {Simmons}, \citenamefont {Ward}, \citenamefont {Prance},
  \citenamefont {Wu}, \citenamefont {Koh}, \citenamefont {Gamble},
  \citenamefont {Savage}, \citenamefont {Lagally}, \citenamefont {Friesen},
  \citenamefont {Coppersmith},\ and\ \citenamefont {Eriksson}}]{shi2014}%
  \BibitemOpen
  \bibfield  {author} {\bibinfo {author} {\bibfnamefont {Z.}~\bibnamefont
  {Shi}}, \bibinfo {author} {\bibfnamefont {C.~B.}\ \bibnamefont {Simmons}},
  \bibinfo {author} {\bibfnamefont {D.~R.}\ \bibnamefont {Ward}}, \bibinfo
  {author} {\bibfnamefont {J.~R.}\ \bibnamefont {Prance}}, \bibinfo {author}
  {\bibfnamefont {X.}~\bibnamefont {Wu}}, \bibinfo {author} {\bibfnamefont
  {T.~S.}\ \bibnamefont {Koh}}, \bibinfo {author} {\bibfnamefont {J.~K.}\
  \bibnamefont {Gamble}}, \bibinfo {author} {\bibfnamefont {D.~E.}\
  \bibnamefont {Savage}}, \bibinfo {author} {\bibfnamefont {M.~G.}\
  \bibnamefont {Lagally}}, \bibinfo {author} {\bibfnamefont {M.}~\bibnamefont
  {Friesen}}, \bibinfo {author} {\bibfnamefont {S.~N.}\ \bibnamefont
  {Coppersmith}}, \ and\ \bibinfo {author} {\bibfnamefont {M.~A.}\ \bibnamefont
  {Eriksson}},\ }\bibfield  {title} {\enquote {\bibinfo {title} {Fast coherent
  manipulation of three-electron states in a double quantum dot},}\ }\href
  {\doibase 10.1038/ncomms4020} {\bibfield  {journal} {\bibinfo  {journal}
  {Nat. Commun.}\ }\textbf {\bibinfo {volume} {5}},\ \bibinfo {pages} {3020}
  (\bibinfo {year} {2014})}\BibitemShut {NoStop}%
\bibitem [{\citenamefont {Kim}\ \emph {et~al.}(2014)\citenamefont {Kim},
  \citenamefont {Shi}, \citenamefont {Simmons}, \citenamefont {Ward},
  \citenamefont {Prance}, \citenamefont {Koh}, \citenamefont {Gamble},
  \citenamefont {Savage}, \citenamefont {Lagally}, \citenamefont {Friesen},
  \citenamefont {Coppersmith},\ and\ \citenamefont {Eriksson}}]{kim2014}%
  \BibitemOpen
  \bibfield  {author} {\bibinfo {author} {\bibfnamefont {D.}~\bibnamefont
  {Kim}}, \bibinfo {author} {\bibfnamefont {Z.}~\bibnamefont {Shi}}, \bibinfo
  {author} {\bibfnamefont {C.~B.}\ \bibnamefont {Simmons}}, \bibinfo {author}
  {\bibfnamefont {D.~R.}\ \bibnamefont {Ward}}, \bibinfo {author}
  {\bibfnamefont {J.~R.}\ \bibnamefont {Prance}}, \bibinfo {author}
  {\bibfnamefont {T.~S.}\ \bibnamefont {Koh}}, \bibinfo {author} {\bibfnamefont
  {J.~K.}\ \bibnamefont {Gamble}}, \bibinfo {author} {\bibfnamefont {D.~E.}\
  \bibnamefont {Savage}}, \bibinfo {author} {\bibfnamefont {M.~G.}\
  \bibnamefont {Lagally}}, \bibinfo {author} {\bibfnamefont {M.}~\bibnamefont
  {Friesen}}, \bibinfo {author} {\bibfnamefont {S.~N.}\ \bibnamefont
  {Coppersmith}}, \ and\ \bibinfo {author} {\bibfnamefont {M.~A.}\ \bibnamefont
  {Eriksson}},\ }\bibfield  {title} {\enquote {\bibinfo {title} {Quantum
  control and process tomography of a semiconductor quantum dot hybrid
  qubit},}\ }\href {\doibase 10.1038/nature13407} {\bibfield  {journal}
  {\bibinfo  {journal} {Nature (London)}\ }\textbf {\bibinfo {volume} {511}},\
  \bibinfo {pages} {70} (\bibinfo {year} {2014})}\BibitemShut {NoStop}%
\bibitem [{\citenamefont {Kim}\ \emph {et~al.}(2015{\natexlab{a}})\citenamefont
  {Kim}, \citenamefont {Ward}, \citenamefont {Simmons}, \citenamefont {Gamble},
  \citenamefont {Blume-Kohout}, \citenamefont {Nielsen}, \citenamefont
  {Savage}, \citenamefont {Lagally}, \citenamefont {Friesen}, \citenamefont
  {Coppersmith},\ and\ \citenamefont {Eriksson}}]{kim2014-2}%
  \BibitemOpen
  \bibfield  {author} {\bibinfo {author} {\bibfnamefont {D.}~\bibnamefont
  {Kim}}, \bibinfo {author} {\bibfnamefont {D.~R.}\ \bibnamefont {Ward}},
  \bibinfo {author} {\bibfnamefont {C.~B.}\ \bibnamefont {Simmons}}, \bibinfo
  {author} {\bibfnamefont {J.~K.}\ \bibnamefont {Gamble}}, \bibinfo {author}
  {\bibfnamefont {R.}~\bibnamefont {Blume-Kohout}}, \bibinfo {author}
  {\bibfnamefont {E.}~\bibnamefont {Nielsen}}, \bibinfo {author} {\bibfnamefont
  {D.~E.}\ \bibnamefont {Savage}}, \bibinfo {author} {\bibfnamefont {M.~G.}\
  \bibnamefont {Lagally}}, \bibinfo {author} {\bibfnamefont {Mark}\
  \bibnamefont {Friesen}}, \bibinfo {author} {\bibfnamefont {S.~N.}\
  \bibnamefont {Coppersmith}}, \ and\ \bibinfo {author} {\bibfnamefont {M.~A.}\
  \bibnamefont {Eriksson}},\ }\bibfield  {title} {\enquote {\bibinfo {title}
  {Microwave-driven coherent operations of a semiconductor quantum dot charge
  qubit},}\ }\href {\doibase 10.1038/nnano.2014.336} {\bibfield  {journal}
  {\bibinfo  {journal} {Nat. Nanotechnol.}\ }\textbf {\bibinfo {volume} {10}},\
  \bibinfo {pages} {243} (\bibinfo {year} {2015}{\natexlab{a}})}\BibitemShut
  {NoStop}%
\bibitem [{\citenamefont {Petersson}\ \emph {et~al.}(2010)\citenamefont
  {Petersson}, \citenamefont {Petta}, \citenamefont {Lu},\ and\ \citenamefont
  {Gossard}}]{petersson2010}%
  \BibitemOpen
  \bibfield  {author} {\bibinfo {author} {\bibfnamefont {K.~D.}\ \bibnamefont
  {Petersson}}, \bibinfo {author} {\bibfnamefont {J.~R.}\ \bibnamefont
  {Petta}}, \bibinfo {author} {\bibfnamefont {H.}~\bibnamefont {Lu}}, \ and\
  \bibinfo {author} {\bibfnamefont {A.~C.}\ \bibnamefont {Gossard}},\
  }\bibfield  {title} {\enquote {\bibinfo {title} {{Quantum Coherence in a
  One-Electron Semiconductor Charge Qubit}},}\ }\href {\doibase
  10.1103/PhysRevLett.105.246804} {\bibfield  {journal} {\bibinfo  {journal}
  {Phys. Rev. Lett.}\ }\textbf {\bibinfo {volume} {105}},\ \bibinfo {pages}
  {246804} (\bibinfo {year} {2010})}\BibitemShut {NoStop}%
\bibitem [{\citenamefont {Dovzhenko}\ \emph {et~al.}(2011)\citenamefont
  {Dovzhenko}, \citenamefont {J.~Stehlik}, \citenamefont {Petta}, \citenamefont
  {Lu},\ and\ \citenamefont {Gossard}}]{dovzhenko2011}%
  \BibitemOpen
  \bibfield  {author} {\bibinfo {author} {\bibfnamefont {Y.}~\bibnamefont
  {Dovzhenko}}, \bibinfo {author} {\bibfnamefont {K.~D.~Petersson}\
  \bibnamefont {J.~Stehlik}}, \bibinfo {author} {\bibfnamefont {J.~R.}\
  \bibnamefont {Petta}}, \bibinfo {author} {\bibfnamefont {H.}~\bibnamefont
  {Lu}}, \ and\ \bibinfo {author} {\bibfnamefont {A.~C.}\ \bibnamefont
  {Gossard}},\ }\bibfield  {title} {\enquote {\bibinfo {title} {Nonadiabatic
  quantum control of a semiconductor charge qubit},}\ }\href {\doibase
  10.1103/PhysRevB.84.161302} {\bibfield  {journal} {\bibinfo  {journal} {Phys.
  Rev. B}\ }\textbf {\bibinfo {volume} {84}},\ \bibinfo {pages} {161302}
  (\bibinfo {year} {2011})}\BibitemShut {NoStop}%
\bibitem [{\citenamefont {Cao}\ \emph {et~al.}(2013)\citenamefont {Cao},
  \citenamefont {Li}, \citenamefont {Tu}, \citenamefont {Wang}, \citenamefont
  {Zhou}, \citenamefont {Xiao}, \citenamefont {Guo}, \citenamefont {Jiang},\
  and\ \citenamefont {Guo}}]{cao2013}%
  \BibitemOpen
  \bibfield  {author} {\bibinfo {author} {\bibfnamefont {G.}~\bibnamefont
  {Cao}}, \bibinfo {author} {\bibfnamefont {H.-O.}\ \bibnamefont {Li}},
  \bibinfo {author} {\bibfnamefont {T.}~\bibnamefont {Tu}}, \bibinfo {author}
  {\bibfnamefont {L.}~\bibnamefont {Wang}}, \bibinfo {author} {\bibfnamefont
  {C.}~\bibnamefont {Zhou}}, \bibinfo {author} {\bibfnamefont {M.}~\bibnamefont
  {Xiao}}, \bibinfo {author} {\bibfnamefont {G.-C.}\ \bibnamefont {Guo}},
  \bibinfo {author} {\bibfnamefont {H.-W.}\ \bibnamefont {Jiang}}, \ and\
  \bibinfo {author} {\bibfnamefont {G.-P.}\ \bibnamefont {Guo}},\ }\bibfield
  {title} {\enquote {\bibinfo {title} {{Ultrafast universal quantum control of
  a quantum-dot charge qubit using Landau--Zener--St{\"u}ckelberg
  interference}},}\ }\href {\doibase 10.1038/ncomms2412} {\bibfield  {journal}
  {\bibinfo  {journal} {Nat. Commun.}\ }\textbf {\bibinfo {volume} {4}},\
  \bibinfo {pages} {1401} (\bibinfo {year} {2013})}\BibitemShut {NoStop}%
\bibitem [{\citenamefont {Vion}\ \emph {et~al.}(2002)\citenamefont {Vion},
  \citenamefont {Aassime}, \citenamefont {Cottet}, \citenamefont {Joyez},
  \citenamefont {Pothier}, \citenamefont {Urbina}, \citenamefont {Esteve},\
  and\ \citenamefont {Devoret}}]{vion2002}%
  \BibitemOpen
  \bibfield  {author} {\bibinfo {author} {\bibfnamefont {D.}~\bibnamefont
  {Vion}}, \bibinfo {author} {\bibfnamefont {A.}~\bibnamefont {Aassime}},
  \bibinfo {author} {\bibfnamefont {A.}~\bibnamefont {Cottet}}, \bibinfo
  {author} {\bibfnamefont {P.}~\bibnamefont {Joyez}}, \bibinfo {author}
  {\bibfnamefont {H.}~\bibnamefont {Pothier}}, \bibinfo {author} {\bibfnamefont
  {C.}~\bibnamefont {Urbina}}, \bibinfo {author} {\bibfnamefont
  {D.}~\bibnamefont {Esteve}}, \ and\ \bibinfo {author} {\bibfnamefont {M.~H.}\
  \bibnamefont {Devoret}},\ }\bibfield  {title} {\enquote {\bibinfo {title}
  {{Manipulating the Quantum State of an Electrical Circuit}},}\ }\href
  {\doibase 10.1126/science.1069372} {\bibfield  {journal} {\bibinfo  {journal}
  {Science}\ }\textbf {\bibinfo {volume} {296}},\ \bibinfo {pages} {886}
  (\bibinfo {year} {2002})}\BibitemShut {NoStop}%
\bibitem [{\citenamefont {Koch}\ \emph {et~al.}(2007)\citenamefont {Koch},
  \citenamefont {Yu}, \citenamefont {Gambetta}, \citenamefont {Houck},
  \citenamefont {Schuster}, \citenamefont {Majer}, \citenamefont {Blais},
  \citenamefont {Devoret}, \citenamefont {Girvin},\ and\ \citenamefont
  {Schoelkopf}}]{koch2007}%
  \BibitemOpen
  \bibfield  {author} {\bibinfo {author} {\bibfnamefont {J.}~\bibnamefont
  {Koch}}, \bibinfo {author} {\bibfnamefont {T.~M.}\ \bibnamefont {Yu}},
  \bibinfo {author} {\bibfnamefont {J.}~\bibnamefont {Gambetta}}, \bibinfo
  {author} {\bibfnamefont {A.~A.}\ \bibnamefont {Houck}}, \bibinfo {author}
  {\bibfnamefont {D.~I.}\ \bibnamefont {Schuster}}, \bibinfo {author}
  {\bibfnamefont {J.}~\bibnamefont {Majer}}, \bibinfo {author} {\bibfnamefont
  {Alexandre}\ \bibnamefont {Blais}}, \bibinfo {author} {\bibfnamefont {M.~H.}\
  \bibnamefont {Devoret}}, \bibinfo {author} {\bibfnamefont {S.~M.}\
  \bibnamefont {Girvin}}, \ and\ \bibinfo {author} {\bibfnamefont {R.~J.}\
  \bibnamefont {Schoelkopf}},\ }\bibfield  {title} {\enquote {\bibinfo {title}
  {{Charge-insensitive qubit design derived from the Cooper pair box}},}\
  }\href {\doibase 10.1103/PhysRevA.76.042319} {\bibfield  {journal} {\bibinfo
  {journal} {Phys. Rev. A}\ }\textbf {\bibinfo {volume} {76}},\ \bibinfo
  {pages} {042319} (\bibinfo {year} {2007})}\BibitemShut {NoStop}%
\bibitem [{\citenamefont {Schreier}\ \emph {et~al.}(2008)\citenamefont
  {Schreier}, \citenamefont {Houck}, \citenamefont {Koch}, \citenamefont
  {Schuster}, \citenamefont {Johnson}, \citenamefont {Chow}, \citenamefont
  {Gambetta}, \citenamefont {Majer}, \citenamefont {Frunzio}, \citenamefont
  {Devoret}, \citenamefont {Girvin},\ and\ \citenamefont
  {Schoelkopf}}]{schreier2008}%
  \BibitemOpen
  \bibfield  {author} {\bibinfo {author} {\bibfnamefont {J.~A.}\ \bibnamefont
  {Schreier}}, \bibinfo {author} {\bibfnamefont {A.~A.}\ \bibnamefont {Houck}},
  \bibinfo {author} {\bibfnamefont {J.}~\bibnamefont {Koch}}, \bibinfo {author}
  {\bibfnamefont {D.~I.}\ \bibnamefont {Schuster}}, \bibinfo {author}
  {\bibfnamefont {B.~R.}\ \bibnamefont {Johnson}}, \bibinfo {author}
  {\bibfnamefont {J.~M.}\ \bibnamefont {Chow}}, \bibinfo {author}
  {\bibfnamefont {J.~M.}\ \bibnamefont {Gambetta}}, \bibinfo {author}
  {\bibfnamefont {J.}~\bibnamefont {Majer}}, \bibinfo {author} {\bibfnamefont
  {L.}~\bibnamefont {Frunzio}}, \bibinfo {author} {\bibfnamefont {M.~H.}\
  \bibnamefont {Devoret}}, \bibinfo {author} {\bibfnamefont {S.~M.}\
  \bibnamefont {Girvin}}, \ and\ \bibinfo {author} {\bibfnamefont {R.~J.}\
  \bibnamefont {Schoelkopf}},\ }\bibfield  {title} {\enquote {\bibinfo {title}
  {Suppressing charge noise decoherence in superconducting charge qubits},}\
  }\href {\doibase 10.1103/PhysRevB.77.180502} {\bibfield  {journal} {\bibinfo
  {journal} {Phys. Rev. B}\ }\textbf {\bibinfo {volume} {77}},\ \bibinfo
  {pages} {180502} (\bibinfo {year} {2008})}\BibitemShut {NoStop}%
\bibitem [{\citenamefont {Shulman}\ \emph {et~al.}(2014)\citenamefont
  {Shulman}, \citenamefont {Harvey}, \citenamefont {Nichol}, \citenamefont
  {Bartlett}, \citenamefont {Doherty}, \citenamefont {Umansky},\ and\
  \citenamefont {Yacoby}}]{shulman2014}%
  \BibitemOpen
  \bibfield  {author} {\bibinfo {author} {\bibfnamefont {M.~D.}\ \bibnamefont
  {Shulman}}, \bibinfo {author} {\bibfnamefont {S.~P.}\ \bibnamefont {Harvey}},
  \bibinfo {author} {\bibfnamefont {J.~M.}\ \bibnamefont {Nichol}}, \bibinfo
  {author} {\bibfnamefont {S.~D.}\ \bibnamefont {Bartlett}}, \bibinfo {author}
  {\bibfnamefont {A.~C.}\ \bibnamefont {Doherty}}, \bibinfo {author}
  {\bibfnamefont {V.}~\bibnamefont {Umansky}}, \ and\ \bibinfo {author}
  {\bibfnamefont {A.}~\bibnamefont {Yacoby}},\ }\bibfield  {title} {\enquote
  {\bibinfo {title} {{Suppressing qubit dephasing using real-time Hamiltonian
  estimation}},}\ }\href {\doibase 10.1038/ncomms6156} {\bibfield  {journal}
  {\bibinfo  {journal} {Nat. Commun.}\ }\textbf {\bibinfo {volume} {5}},\
  \bibinfo {pages} {5156} (\bibinfo {year} {2014})}\BibitemShut {NoStop}%
\bibitem [{\citenamefont {Taylor}\ \emph {et~al.}(2013)\citenamefont {Taylor},
  \citenamefont {Srinivasa},\ and\ \citenamefont {Medford}}]{taylor2013}%
  \BibitemOpen
  \bibfield  {author} {\bibinfo {author} {\bibfnamefont {J.~M.}\ \bibnamefont
  {Taylor}}, \bibinfo {author} {\bibfnamefont {V.}~\bibnamefont {Srinivasa}}, \
  and\ \bibinfo {author} {\bibfnamefont {J.}~\bibnamefont {Medford}},\
  }\bibfield  {title} {\enquote {\bibinfo {title} {{Electrically Protected
  Resonant Exchange Qubits in Triple Quantum Dots}},}\ }\href {\doibase
  10.1103/PhysRevLett.111.050502} {\bibfield  {journal} {\bibinfo  {journal}
  {Phys. Rev. Lett.}\ }\textbf {\bibinfo {volume} {111}},\ \bibinfo {pages}
  {050502} (\bibinfo {year} {2013})}\BibitemShut {NoStop}%
\bibitem [{\citenamefont {Medford}\ \emph
  {et~al.}(2013{\natexlab{a}})\citenamefont {Medford}, \citenamefont {Beil},
  \citenamefont {Taylor}, \citenamefont {Rashba}, \citenamefont {Lu},
  \citenamefont {Gossard},\ and\ \citenamefont {Marcus}}]{medford2013}%
  \BibitemOpen
  \bibfield  {author} {\bibinfo {author} {\bibfnamefont {J.}~\bibnamefont
  {Medford}}, \bibinfo {author} {\bibfnamefont {J.}~\bibnamefont {Beil}},
  \bibinfo {author} {\bibfnamefont {J.~M.}\ \bibnamefont {Taylor}}, \bibinfo
  {author} {\bibfnamefont {E.~I.}\ \bibnamefont {Rashba}}, \bibinfo {author}
  {\bibfnamefont {H.}~\bibnamefont {Lu}}, \bibinfo {author} {\bibfnamefont
  {A.~C.}\ \bibnamefont {Gossard}}, \ and\ \bibinfo {author} {\bibfnamefont
  {C.~M.}\ \bibnamefont {Marcus}},\ }\bibfield  {title} {\enquote {\bibinfo
  {title} {{Quantum-Dot-Based Resonant Exchange Qubit}},}\ }\href {\doibase
  10.1103/PhysRevLett.111.050501} {\bibfield  {journal} {\bibinfo  {journal}
  {Phys. Rev. Lett.}\ }\textbf {\bibinfo {volume} {111}},\ \bibinfo {pages}
  {050501} (\bibinfo {year} {2013}{\natexlab{a}})}\BibitemShut {NoStop}%
\bibitem [{\citenamefont {Mehl}(2015)}]{mehl2015-1}%
  \BibitemOpen
  \bibfield  {author} {\bibinfo {author} {\bibfnamefont {S.}~\bibnamefont
  {Mehl}},\ }\bibfield  {title} {\enquote {\bibinfo {title} {Two-qubit pulse
  gate for the three-electron double quantum dot qubit},}\ }\href {\doibase
  10.1103/PhysRevB.91.035430} {\bibfield  {journal} {\bibinfo  {journal} {Phys.
  Rev. B}\ }\textbf {\bibinfo {volume} {91}},\ \bibinfo {pages} {035430}
  (\bibinfo {year} {2015})}\BibitemShut {NoStop}%
\bibitem [{\citenamefont {Mehl}\ and\ \citenamefont
  {DiVincenzo}(2013{\natexlab{a}})}]{mehl2013-1}%
  \BibitemOpen
  \bibfield  {author} {\bibinfo {author} {\bibfnamefont {S.}~\bibnamefont
  {Mehl}}\ and\ \bibinfo {author} {\bibfnamefont {D.~P.}\ \bibnamefont
  {DiVincenzo}},\ }\bibfield  {title} {\enquote {\bibinfo {title} {{Noise
  analysis of qubits implemented in triple quantum dot systems in a Davies
  master equation approach}},}\ }\href {\doibase 10.1103/PhysRevB.87.195309}
  {\bibfield  {journal} {\bibinfo  {journal} {Phys. Rev. B}\ }\textbf {\bibinfo
  {volume} {87}},\ \bibinfo {pages} {195309} (\bibinfo {year}
  {2013}{\natexlab{a}})}\BibitemShut {NoStop}%
\bibitem [{Note1()}]{Note1}%
  \BibitemOpen
  \bibinfo {note} {Note that the tunnel couplings between HQs couple between
  spin subspaces such that two-qubit gates can be problematic if the two HQs
  are initialized to different spin subspaces.}\BibitemShut {Stop}%
\bibitem [{Note2()}]{Note2}%
  \BibitemOpen
  \bibinfo {note} {I define that \protect {$\delimiter "3222378 $} is the spin
  configuration with the lowest energy. Note that there are materials of
  positive g factors (e.g. Si) and negative g factors (e.g. GaAs).}\BibitemShut
  {Stop}%
\bibitem [{\citenamefont {Sakurai}\ and\ \citenamefont
  {Tuan}(1994)}]{sakurai1994}%
  \BibitemOpen
  \bibfield  {author} {\bibinfo {author} {\bibfnamefont {J.~J.}\ \bibnamefont
  {Sakurai}}\ and\ \bibinfo {author} {\bibfnamefont {S.~F.}\ \bibnamefont
  {Tuan}},\ }\href@noop {} {\emph {\bibinfo {title} {Modern Quantum
  Mechanics}}}\ (\bibinfo  {publisher} {Addison-Wesley},\ \bibinfo {address}
  {Reading},\ \bibinfo {year} {1994})\BibitemShut {NoStop}%
\bibitem [{\citenamefont {Reilly}\ \emph {et~al.}(2007)\citenamefont {Reilly},
  \citenamefont {Marcus}, \citenamefont {Hanson},\ and\ \citenamefont
  {Gossard}}]{reilly2007}%
  \BibitemOpen
  \bibfield  {author} {\bibinfo {author} {\bibfnamefont {D.~J.}\ \bibnamefont
  {Reilly}}, \bibinfo {author} {\bibfnamefont {C.~M.}\ \bibnamefont {Marcus}},
  \bibinfo {author} {\bibfnamefont {M.~P.}\ \bibnamefont {Hanson}}, \ and\
  \bibinfo {author} {\bibfnamefont {A.~C.}\ \bibnamefont {Gossard}},\
  }\bibfield  {title} {\enquote {\bibinfo {title} {Fast single-charge sensing
  with a rf quantum point contact},}\ }\href {\doibase 10.1063/1.2794995}
  {\bibfield  {journal} {\bibinfo  {journal} {Appl. Phys. Lett.}\ }\textbf
  {\bibinfo {volume} {91}},\ \bibinfo {pages} {162101} (\bibinfo {year}
  {2007})}\BibitemShut {NoStop}%
\bibitem [{\citenamefont {Barthel}\ \emph {et~al.}(2009)\citenamefont
  {Barthel}, \citenamefont {Reilly}, \citenamefont {Marcus}, \citenamefont
  {Hanson},\ and\ \citenamefont {Gossard}}]{barthel2009}%
  \BibitemOpen
  \bibfield  {author} {\bibinfo {author} {\bibfnamefont {C.}~\bibnamefont
  {Barthel}}, \bibinfo {author} {\bibfnamefont {D.~J.}\ \bibnamefont {Reilly}},
  \bibinfo {author} {\bibfnamefont {C.~M.}\ \bibnamefont {Marcus}}, \bibinfo
  {author} {\bibfnamefont {M.~P.}\ \bibnamefont {Hanson}}, \ and\ \bibinfo
  {author} {\bibfnamefont {A.~C.}\ \bibnamefont {Gossard}},\ }\bibfield
  {title} {\enquote {\bibinfo {title} {{Rapid Single-Shot Measurement of a
  Singlet-Triplet Qubit}},}\ }\href {\doibase 10.1103/PhysRevLett.103.160503}
  {\bibfield  {journal} {\bibinfo  {journal} {Phys. Rev. Lett.}\ }\textbf
  {\bibinfo {volume} {103}},\ \bibinfo {pages} {160503} (\bibinfo {year}
  {2009})}\BibitemShut {NoStop}%
\bibitem [{\citenamefont {Barthel}\ \emph
  {et~al.}(2010{\natexlab{a}})\citenamefont {Barthel}, \citenamefont
  {Kj{\ae}rgaard}, \citenamefont {Medford}, \citenamefont {Stopa},
  \citenamefont {Marcus}, \citenamefont {Hanson},\ and\ \citenamefont
  {Gossard}}]{barthel2010-2}%
  \BibitemOpen
  \bibfield  {author} {\bibinfo {author} {\bibfnamefont {C.}~\bibnamefont
  {Barthel}}, \bibinfo {author} {\bibfnamefont {M.}~\bibnamefont
  {Kj{\ae}rgaard}}, \bibinfo {author} {\bibfnamefont {J.}~\bibnamefont
  {Medford}}, \bibinfo {author} {\bibfnamefont {M.}~\bibnamefont {Stopa}},
  \bibinfo {author} {\bibfnamefont {C.~M.}\ \bibnamefont {Marcus}}, \bibinfo
  {author} {\bibfnamefont {M.~P.}\ \bibnamefont {Hanson}}, \ and\ \bibinfo
  {author} {\bibfnamefont {A.~C.}\ \bibnamefont {Gossard}},\ }\bibfield
  {title} {\enquote {\bibinfo {title} {Fast sensing of double-dot charge
  arrangement and spin state with a radio-frequency sensor quantum dot},}\
  }\href {\doibase 10.1103/PhysRevB.81.161308} {\bibfield  {journal} {\bibinfo
  {journal} {Phys. Rev. B}\ }\textbf {\bibinfo {volume} {81}},\ \bibinfo
  {pages} {161308} (\bibinfo {year} {2010}{\natexlab{a}})}\BibitemShut
  {NoStop}%
\bibitem [{\citenamefont {Hu}\ and\ \citenamefont {{Das
  Sarma}}(2006)}]{hu2006}%
  \BibitemOpen
  \bibfield  {author} {\bibinfo {author} {\bibfnamefont {X.}~\bibnamefont
  {Hu}}\ and\ \bibinfo {author} {\bibfnamefont {S.}~\bibnamefont {{Das
  Sarma}}},\ }\bibfield  {title} {\enquote {\bibinfo {title}
  {{Charge-Fluctuation-Induced Dephasing of Exchange-Coupled Spin Qubits}},}\
  }\href {\doibase 10.1103/PhysRevLett.96.100501} {\bibfield  {journal}
  {\bibinfo  {journal} {Phys. Rev. Lett.}\ }\textbf {\bibinfo {volume} {96}},\
  \bibinfo {pages} {100501} (\bibinfo {year} {2006})}\BibitemShut {NoStop}%
\bibitem [{\citenamefont {Dial}\ \emph {et~al.}(2013)\citenamefont {Dial},
  \citenamefont {Shulman}, \citenamefont {Harvey}, \citenamefont {Bluhm},
  \citenamefont {Umansky},\ and\ \citenamefont {Yacoby}}]{dial2013}%
  \BibitemOpen
  \bibfield  {author} {\bibinfo {author} {\bibfnamefont {O.~E.}\ \bibnamefont
  {Dial}}, \bibinfo {author} {\bibfnamefont {M.~D.}\ \bibnamefont {Shulman}},
  \bibinfo {author} {\bibfnamefont {S.~P.}\ \bibnamefont {Harvey}}, \bibinfo
  {author} {\bibfnamefont {H.}~\bibnamefont {Bluhm}}, \bibinfo {author}
  {\bibfnamefont {V.}~\bibnamefont {Umansky}}, \ and\ \bibinfo {author}
  {\bibfnamefont {A.}~\bibnamefont {Yacoby}},\ }\bibfield  {title} {\enquote
  {\bibinfo {title} {{Charge Noise Spectroscopy Using Coherent Exchange
  Oscillations in a Singlet-Triplet Qubit}},}\ }\href {\doibase
  10.1103/PhysRevLett.110.146804} {\bibfield  {journal} {\bibinfo  {journal}
  {Phys. Rev. Lett.}\ }\textbf {\bibinfo {volume} {110}},\ \bibinfo {pages}
  {146804} (\bibinfo {year} {2013})}\BibitemShut {NoStop}%
\bibitem [{\citenamefont {Yang}\ \emph {et~al.}(2013)\citenamefont {Yang},
  \citenamefont {Rossi}, \citenamefont {Ruskov}, \citenamefont {Lai},
  \citenamefont {Mohiyaddin}, \citenamefont {Lee}, \citenamefont {Tahan},
  \citenamefont {Klimeck}, \citenamefont {Morello},\ and\ \citenamefont
  {Dzurak}}]{yang2013}%
  \BibitemOpen
  \bibfield  {author} {\bibinfo {author} {\bibfnamefont {C.~H.}\ \bibnamefont
  {Yang}}, \bibinfo {author} {\bibfnamefont {A.}~\bibnamefont {Rossi}},
  \bibinfo {author} {\bibfnamefont {R.}~\bibnamefont {Ruskov}}, \bibinfo
  {author} {\bibfnamefont {N.~S.}\ \bibnamefont {Lai}}, \bibinfo {author}
  {\bibfnamefont {F.~A.}\ \bibnamefont {Mohiyaddin}}, \bibinfo {author}
  {\bibfnamefont {S.}~\bibnamefont {Lee}}, \bibinfo {author} {\bibfnamefont
  {C.}~\bibnamefont {Tahan}}, \bibinfo {author} {\bibfnamefont
  {G.}~\bibnamefont {Klimeck}}, \bibinfo {author} {\bibfnamefont
  {A.}~\bibnamefont {Morello}}, \ and\ \bibinfo {author} {\bibfnamefont
  {A.~S.}\ \bibnamefont {Dzurak}},\ }\bibfield  {title} {\enquote {\bibinfo
  {title} {Spin-valley lifetimes in a silicon quantum dot with tunable valley
  splitting},}\ }\href@noop {} {\bibfield  {journal} {\bibinfo  {journal} {Nat.
  Commun.}\ }\textbf {\bibinfo {volume} {4}},\ \bibinfo {pages} {2069}
  (\bibinfo {year} {2013})}\BibitemShut {NoStop}%
\bibitem [{\citenamefont {Mehl}\ and\ \citenamefont
  {DiVincenzo}(2013{\natexlab{b}})}]{mehl2013-2}%
  \BibitemOpen
  \bibfield  {author} {\bibinfo {author} {\bibfnamefont {S.}~\bibnamefont
  {Mehl}}\ and\ \bibinfo {author} {\bibfnamefont {D.~P.}\ \bibnamefont
  {DiVincenzo}},\ }\bibfield  {title} {\enquote {\bibinfo {title}
  {Noise-protected gate for six-electron double-dot qubit},}\ }\href {\doibase
  10.1103/PhysRevB.88.161408} {\bibfield  {journal} {\bibinfo  {journal} {Phys.
  Rev. B}\ }\textbf {\bibinfo {volume} {88}},\ \bibinfo {pages} {161408}
  (\bibinfo {year} {2013}{\natexlab{b}})}\BibitemShut {NoStop}%
\bibitem [{\citenamefont {Mehl}\ and\ \citenamefont
  {DiVincenzo}(2014)}]{mehl2014-2}%
  \BibitemOpen
  \bibfield  {author} {\bibinfo {author} {\bibfnamefont {S.}~\bibnamefont
  {Mehl}}\ and\ \bibinfo {author} {\bibfnamefont {David~P.}\ \bibnamefont
  {DiVincenzo}},\ }\bibfield  {title} {\enquote {\bibinfo {title} {Inverted
  singlet-triplet qubit coded on a two-electron double quantum dot},}\ }\href
  {\doibase 10.1103/PhysRevB.90.195424} {\bibfield  {journal} {\bibinfo
  {journal} {Phys. Rev. B}\ }\textbf {\bibinfo {volume} {90}},\ \bibinfo
  {pages} {195424} (\bibinfo {year} {2014})}\BibitemShut {NoStop}%
\bibitem [{\citenamefont {Hiltunen}\ \emph {et~al.}(2015)\citenamefont
  {Hiltunen}, \citenamefont {Bluhm}, \citenamefont {Mehl},\ and\ \citenamefont
  {Harju}}]{hiltunen2015}%
  \BibitemOpen
  \bibfield  {author} {\bibinfo {author} {\bibfnamefont {T.}~\bibnamefont
  {Hiltunen}}, \bibinfo {author} {\bibfnamefont {H.}~\bibnamefont {Bluhm}},
  \bibinfo {author} {\bibfnamefont {S.}~\bibnamefont {Mehl}}, \ and\ \bibinfo
  {author} {\bibfnamefont {A.}~\bibnamefont {Harju}},\ }\bibfield  {title}
  {\enquote {\bibinfo {title} {Charge-noise tolerant exchange gates of
  singlet-triplet qubits in asymmetric double quantum dots},}\ }\href {\doibase
  10.1103/PhysRevB.91.075301} {\bibfield  {journal} {\bibinfo  {journal} {Phys.
  Rev. B}\ }\textbf {\bibinfo {volume} {91}},\ \bibinfo {pages} {075301}
  (\bibinfo {year} {2015})}\BibitemShut {NoStop}%
\bibitem [{\citenamefont {Slichter}(1990)}]{slichter1990}%
  \BibitemOpen
  \bibfield  {author} {\bibinfo {author} {\bibfnamefont {C.~P.}\ \bibnamefont
  {Slichter}},\ }\href@noop {} {\emph {\bibinfo {title} {Principles of Magnetic
  Resonance}}}\ (\bibinfo  {publisher} {Springer},\ \bibinfo {address}
  {Berlin},\ \bibinfo {year} {1990})\BibitemShut {NoStop}%
\bibitem [{\citenamefont {Vandersypen}\ and\ \citenamefont
  {Chuang}(2005)}]{vandersypen2005}%
  \BibitemOpen
  \bibfield  {author} {\bibinfo {author} {\bibfnamefont {L.~M.~K.}\
  \bibnamefont {Vandersypen}}\ and\ \bibinfo {author} {\bibfnamefont {I.~L.}\
  \bibnamefont {Chuang}},\ }\bibfield  {title} {\enquote {\bibinfo {title}
  {{NMR techniques for quantum control and computation}},}\ }\href {\doibase
  10.1103/RevModPhys.76.1037} {\bibfield  {journal} {\bibinfo  {journal} {Rev.
  Mod. Phys.}\ }\textbf {\bibinfo {volume} {76}},\ \bibinfo {pages} {1037}
  (\bibinfo {year} {2005})}\BibitemShut {NoStop}%
\bibitem [{\citenamefont {{Ioffe Institute}}(2001)}]{ioffe}%
  \BibitemOpen
  \bibfield  {author} {\bibinfo {author} {\bibnamefont {{Ioffe Institute}}},\
  }\href {http://www.ioffe.ru/SVA/NSM/} {\emph {\bibinfo {title} {Electronic
  archive: New Semiconductor Materials. Characteristics and Properties}}}\
  (\bibinfo  {publisher} {http://www.ioffe.ru/SVA/NSM/},\ \bibinfo {year}
  {2001})\BibitemShut {NoStop}%
\bibitem [{\citenamefont {Koh}\ \emph {et~al.}(2012)\citenamefont {Koh},
  \citenamefont {Gamble}, \citenamefont {Friesen}, \citenamefont {Eriksson},\
  and\ \citenamefont {Coppersmith}}]{koh2012}%
  \BibitemOpen
  \bibfield  {author} {\bibinfo {author} {\bibfnamefont {T.~S.}\ \bibnamefont
  {Koh}}, \bibinfo {author} {\bibfnamefont {J.~K.}\ \bibnamefont {Gamble}},
  \bibinfo {author} {\bibfnamefont {M.}~\bibnamefont {Friesen}}, \bibinfo
  {author} {\bibfnamefont {M.~A.}\ \bibnamefont {Eriksson}}, \ and\ \bibinfo
  {author} {\bibfnamefont {S.~N.}\ \bibnamefont {Coppersmith}},\ }\bibfield
  {title} {\enquote {\bibinfo {title} {{Pulse-Gated Quantum-Dot Hybrid
  Qubit}},}\ }\href {\doibase 10.1103/PhysRevLett.109.250503} {\bibfield
  {journal} {\bibinfo  {journal} {Phys. Rev. Lett.}\ }\textbf {\bibinfo
  {volume} {109}},\ \bibinfo {pages} {250503} (\bibinfo {year}
  {2012})}\BibitemShut {NoStop}%
\bibitem [{\citenamefont {Trifunovic}\ \emph {et~al.}(2012)\citenamefont
  {Trifunovic}, \citenamefont {Dial}, \citenamefont {Trif}, \citenamefont
  {Wootton}, \citenamefont {Abebe}, \citenamefont {Yacoby},\ and\ \citenamefont
  {Loss}}]{trifunovic2012}%
  \BibitemOpen
  \bibfield  {author} {\bibinfo {author} {\bibfnamefont {L.}~\bibnamefont
  {Trifunovic}}, \bibinfo {author} {\bibfnamefont {O.}~\bibnamefont {Dial}},
  \bibinfo {author} {\bibfnamefont {M.}~\bibnamefont {Trif}}, \bibinfo {author}
  {\bibfnamefont {J.~R.}\ \bibnamefont {Wootton}}, \bibinfo {author}
  {\bibfnamefont {R.}~\bibnamefont {Abebe}}, \bibinfo {author} {\bibfnamefont
  {A.}~\bibnamefont {Yacoby}}, \ and\ \bibinfo {author} {\bibfnamefont
  {D.}~\bibnamefont {Loss}},\ }\bibfield  {title} {\enquote {\bibinfo {title}
  {{Long-Distance Spin-Spin Coupling via Floating Gates}},}\ }\href {\doibase
  10.1103/PhysRevX.2.011006} {\bibfield  {journal} {\bibinfo  {journal} {Phys.
  Rev. X}\ }\textbf {\bibinfo {volume} {2}},\ \bibinfo {pages} {011006}
  (\bibinfo {year} {2012})}\BibitemShut {NoStop}%
\bibitem [{\citenamefont {Srinivasa}\ and\ \citenamefont
  {Taylor}(2014)}]{srinivasa2014}%
  \BibitemOpen
  \bibfield  {author} {\bibinfo {author} {\bibfnamefont {V.}~\bibnamefont
  {Srinivasa}}\ and\ \bibinfo {author} {\bibfnamefont {J.~M.}\ \bibnamefont
  {Taylor}},\ }\bibfield  {title} {\enquote {\bibinfo {title} {Capacitively
  coupled singlet-triplet qubits in the double charge resonant regime},}\
  }\href {http://de.arxiv.org/abs/1408.4740} {\bibfield  {journal} {\bibinfo
  {journal} {arXiv:1408.4740 [cond-mat.mes-hall]}\ } (\bibinfo {year}
  {2014})}\BibitemShut {NoStop}%
\bibitem [{\citenamefont {Mehl}\ \emph {et~al.}(2015)\citenamefont {Mehl},
  \citenamefont {Bluhm},\ and\ \citenamefont {DiVincenzo}}]{mehl2015-2}%
  \BibitemOpen
  \bibfield  {author} {\bibinfo {author} {\bibfnamefont {S.}~\bibnamefont
  {Mehl}}, \bibinfo {author} {\bibfnamefont {H.}~\bibnamefont {Bluhm}}, \ and\
  \bibinfo {author} {\bibfnamefont {D.~P.}\ \bibnamefont {DiVincenzo}},\
  }\bibfield  {title} {\enquote {\bibinfo {title} {Fault-tolerant quantum
  computation for singlet-triplet qubits with leakage errors},}\ }\href
  {\doibase 10.1103/PhysRevB.91.085419} {\bibfield  {journal} {\bibinfo
  {journal} {Phys. Rev. B}\ }\textbf {\bibinfo {volume} {91}},\ \bibinfo
  {pages} {085419} (\bibinfo {year} {2015})}\BibitemShut {NoStop}%
\bibitem [{\citenamefont {Fong}\ and\ \citenamefont
  {Wandzura}(2011)}]{fong2011}%
  \BibitemOpen
  \bibfield  {author} {\bibinfo {author} {\bibfnamefont {B.~H.}\ \bibnamefont
  {Fong}}\ and\ \bibinfo {author} {\bibfnamefont {S.~M.}\ \bibnamefont
  {Wandzura}},\ }\bibfield  {title} {\enquote {\bibinfo {title} {{Universal
  quantum computation and leakage reduction in the 3-Qubit decoherence free
  subsystem}},}\ }\href@noop {} {\bibfield  {journal} {\bibinfo  {journal}
  {Quantum Inf. Comput.}\ }\textbf {\bibinfo {volume} {11}},\ \bibinfo {pages}
  {1003} (\bibinfo {year} {2011})}\BibitemShut {NoStop}%
\bibitem [{\citenamefont {Weinstein}\ and\ \citenamefont
  {Hellberg}(2005)}]{weinstein2005}%
  \BibitemOpen
  \bibfield  {author} {\bibinfo {author} {\bibfnamefont {Y.~S.}\ \bibnamefont
  {Weinstein}}\ and\ \bibinfo {author} {\bibfnamefont {C.~S.}\ \bibnamefont
  {Hellberg}},\ }\bibfield  {title} {\enquote {\bibinfo {title} {Energetic
  suppression of decoherence in exchange-only quantum computation},}\ }\href
  {\doibase 10.1103/PhysRevA.72.022319} {\bibfield  {journal} {\bibinfo
  {journal} {Phys. Rev. A}\ }\textbf {\bibinfo {volume} {72}},\ \bibinfo
  {pages} {022319} (\bibinfo {year} {2005})}\BibitemShut {NoStop}%
\bibitem [{\citenamefont {Doherty}\ and\ \citenamefont
  {Wardrop}(2013)}]{doherty2013}%
  \BibitemOpen
  \bibfield  {author} {\bibinfo {author} {\bibfnamefont {A.~C.}\ \bibnamefont
  {Doherty}}\ and\ \bibinfo {author} {\bibfnamefont {M.~P.}\ \bibnamefont
  {Wardrop}},\ }\bibfield  {title} {\enquote {\bibinfo {title} {{Two-Qubit
  Gates for Resonant Exchange Qubits}},}\ }\href {\doibase
  10.1103/PhysRevLett.111.050503} {\bibfield  {journal} {\bibinfo  {journal}
  {Phys. Rev. Lett.}\ }\textbf {\bibinfo {volume} {111}},\ \bibinfo {pages}
  {050503} (\bibinfo {year} {2013})}\BibitemShut {NoStop}%
\bibitem [{\citenamefont {Schrieffer}\ and\ \citenamefont
  {Wolff}(1966)}]{schrieffer1966}%
  \BibitemOpen
  \bibfield  {author} {\bibinfo {author} {\bibfnamefont {J.~R.}\ \bibnamefont
  {Schrieffer}}\ and\ \bibinfo {author} {\bibfnamefont {P.~A.}\ \bibnamefont
  {Wolff}},\ }\bibfield  {title} {\enquote {\bibinfo {title} {{Relation between
  the Anderson and Kondo Hamiltonians}},}\ }\href {\doibase
  10.1103/PhysRev.149.491} {\bibfield  {journal} {\bibinfo  {journal} {Phys.
  Rev.}\ }\textbf {\bibinfo {volume} {149}},\ \bibinfo {pages} {491} (\bibinfo
  {year} {1966})}\BibitemShut {NoStop}%
\bibitem [{\citenamefont {Winkler}(2010)}]{winkler2010}%
  \BibitemOpen
  \bibfield  {author} {\bibinfo {author} {\bibfnamefont {R.}~\bibnamefont
  {Winkler}},\ }\href@noop {} {\emph {\bibinfo {title} {Spin--Orbit Coupling
  Effects in Two-Dimensional Electron and Hole Systems}}},\ Springer Tracts in
  Modern Physics Vol. 191\ (\bibinfo  {publisher} {Springer},\ \bibinfo
  {address} {Berlin},\ \bibinfo {year} {2010})\BibitemShut {NoStop}%
\bibitem [{\citenamefont {Bravyi}\ \emph {et~al.}(2011)\citenamefont {Bravyi},
  \citenamefont {DiVincenzo},\ and\ \citenamefont {Loss}}]{bravyi2011}%
  \BibitemOpen
  \bibfield  {author} {\bibinfo {author} {\bibfnamefont {S.}~\bibnamefont
  {Bravyi}}, \bibinfo {author} {\bibfnamefont {D.~P.}\ \bibnamefont
  {DiVincenzo}}, \ and\ \bibinfo {author} {\bibfnamefont {D.}~\bibnamefont
  {Loss}},\ }\bibfield  {title} {\enquote {\bibinfo {title}
  {{Schrieffer---Wolff transformation for quantum many-body systems}},}\ }\href
  {\doibase 10.1016/j.aop.2011.06.004} {\bibfield  {journal} {\bibinfo
  {journal} {Ann. Phys. (N.Y.)}\ }\textbf {\bibinfo {volume} {326}},\ \bibinfo
  {pages} {2793} (\bibinfo {year} {2011})}\BibitemShut {NoStop}%
\bibitem [{Note3()}]{Note3}%
  \BibitemOpen
  \bibinfo {note} {The CPHASE gate can be obtained using the sequence \begin
  {align*} e^{-i\protect \frac {\pi }{4}\sigma _z^{\protect \text {R}}}&
  e^{-i\protect \frac {\pi }{4}\sigma _x^{\protect \text {R}}} \protect \text
  {iSWAP} e^{i\protect \frac {\pi }{4}\sigma _z^{\protect \text {L}}} \protect
  \text {iSWAP} e^{-i\protect \frac {\pi }{4}\sigma _z^{\protect \text {L}}}
  e^{-i\protect \frac {\pi }{4}\sigma _x^{\protect \text {R}}}\\&
  =e^{-i\protect \frac {3\pi }{4}}\protect \text {CPHASE}, \end {align*} such
  that two iSWAPs construct one CPHASE.}\BibitemShut {Stop}%
\bibitem [{\citenamefont {Schuch}\ and\ \citenamefont
  {Siewert}(2003)}]{schuch2003}%
  \BibitemOpen
  \bibfield  {author} {\bibinfo {author} {\bibfnamefont {N.}~\bibnamefont
  {Schuch}}\ and\ \bibinfo {author} {\bibfnamefont {J.}~\bibnamefont
  {Siewert}},\ }\bibfield  {title} {\enquote {\bibinfo {title} {{Natural
  two-qubit gate for quantum computation using the XY interaction}},}\ }\href
  {\doibase 10.1103/PhysRevA.67.032301} {\bibfield  {journal} {\bibinfo
  {journal} {Phys. Rev. A}\ }\textbf {\bibinfo {volume} {67}},\ \bibinfo
  {pages} {032301} (\bibinfo {year} {2003})}\BibitemShut {NoStop}%
\bibitem [{\citenamefont {Makhlin}(2002)}]{makhlin2002}%
  \BibitemOpen
  \bibfield  {author} {\bibinfo {author} {\bibfnamefont {Y.}~\bibnamefont
  {Makhlin}},\ }\bibfield  {title} {\enquote {\bibinfo {title} {{Nonlocal
  Properties of Two-Qubit Gates and Mixed States, and the Optimization of
  Quantum Computations}},}\ }\href {\doibase 10.1023/A:1022144002391}
  {\bibfield  {journal} {\bibinfo  {journal} {Quantum Inf. Process.}\ }\textbf
  {\bibinfo {volume} {1}},\ \bibinfo {pages} {243} (\bibinfo {year}
  {2002})}\BibitemShut {NoStop}%
\bibitem [{\citenamefont {Paraoanu}(2006)}]{paraoanu2006}%
  \BibitemOpen
  \bibfield  {author} {\bibinfo {author} {\bibfnamefont {G.~S.}\ \bibnamefont
  {Paraoanu}},\ }\bibfield  {title} {\enquote {\bibinfo {title}
  {Microwave-induced coupling of superconducting qubits},}\ }\href {\doibase
  10.1103/PhysRevB.74.140504} {\bibfield  {journal} {\bibinfo  {journal} {Phys.
  Rev. B}\ }\textbf {\bibinfo {volume} {74}},\ \bibinfo {pages} {140504}
  (\bibinfo {year} {2006})}\BibitemShut {NoStop}%
\bibitem [{\citenamefont {Rigetti}\ and\ \citenamefont
  {Devoret}(2010)}]{rigetti2010}%
  \BibitemOpen
  \bibfield  {author} {\bibinfo {author} {\bibfnamefont {C.}~\bibnamefont
  {Rigetti}}\ and\ \bibinfo {author} {\bibfnamefont {M.}~\bibnamefont
  {Devoret}},\ }\bibfield  {title} {\enquote {\bibinfo {title} {Fully
  microwave-tunable universal gates in superconducting qubits with linear
  couplings and fixed transition frequencies},}\ }\href {\doibase
  10.1103/PhysRevB.81.134507} {\bibfield  {journal} {\bibinfo  {journal} {Phys.
  Rev. B}\ }\textbf {\bibinfo {volume} {81}},\ \bibinfo {pages} {134507}
  (\bibinfo {year} {2010})}\BibitemShut {NoStop}%
\bibitem [{\citenamefont {Chow}\ \emph {et~al.}(2011)\citenamefont {Chow},
  \citenamefont {C\'orcoles}, \citenamefont {Gambetta}, \citenamefont
  {Rigetti}, \citenamefont {Johnson}, \citenamefont {Smolin}, \citenamefont
  {Rozen}, \citenamefont {Keefe}, \citenamefont {Rothwell}, \citenamefont
  {Ketchen},\ and\ \citenamefont {Steffen}}]{chow2011}%
  \BibitemOpen
  \bibfield  {author} {\bibinfo {author} {\bibfnamefont {J.~M.}\ \bibnamefont
  {Chow}}, \bibinfo {author} {\bibfnamefont {A.~D.}\ \bibnamefont
  {C\'orcoles}}, \bibinfo {author} {\bibfnamefont {J.~M.}\ \bibnamefont
  {Gambetta}}, \bibinfo {author} {\bibfnamefont {C.}~\bibnamefont {Rigetti}},
  \bibinfo {author} {\bibfnamefont {B.~R.}\ \bibnamefont {Johnson}}, \bibinfo
  {author} {\bibfnamefont {J.~A.}\ \bibnamefont {Smolin}}, \bibinfo {author}
  {\bibfnamefont {J.~R.}\ \bibnamefont {Rozen}}, \bibinfo {author}
  {\bibfnamefont {G.~A.}\ \bibnamefont {Keefe}}, \bibinfo {author}
  {\bibfnamefont {M.~B.}\ \bibnamefont {Rothwell}}, \bibinfo {author}
  {\bibfnamefont {M.~B.}\ \bibnamefont {Ketchen}}, \ and\ \bibinfo {author}
  {\bibfnamefont {M.}~\bibnamefont {Steffen}},\ }\bibfield  {title} {\enquote
  {\bibinfo {title} {{Simple All-Microwave Entangling Gate for Fixed-Frequency
  Superconducting Qubits}},}\ }\href {\doibase 10.1103/PhysRevLett.107.080502}
  {\bibfield  {journal} {\bibinfo  {journal} {Phys. Rev. Lett.}\ }\textbf
  {\bibinfo {volume} {107}},\ \bibinfo {pages} {080502} (\bibinfo {year}
  {2011})}\BibitemShut {NoStop}%
\bibitem [{\citenamefont {Chow}\ \emph {et~al.}(2012)\citenamefont {Chow},
  \citenamefont {Gambetta}, \citenamefont {C\'orcoles}, \citenamefont {Merkel},
  \citenamefont {Smolin}, \citenamefont {Rigetti}, \citenamefont {Poletto},
  \citenamefont {Keefe}, \citenamefont {Rothwell}, \citenamefont {Rozen},
  \citenamefont {Ketchen},\ and\ \citenamefont {Steffen}}]{chow2012}%
  \BibitemOpen
  \bibfield  {author} {\bibinfo {author} {\bibfnamefont {J.~M.}\ \bibnamefont
  {Chow}}, \bibinfo {author} {\bibfnamefont {J.~M.}\ \bibnamefont {Gambetta}},
  \bibinfo {author} {\bibfnamefont {A.~D.}\ \bibnamefont {C\'orcoles}},
  \bibinfo {author} {\bibfnamefont {S.~T.}\ \bibnamefont {Merkel}}, \bibinfo
  {author} {\bibfnamefont {J.~A.}\ \bibnamefont {Smolin}}, \bibinfo {author}
  {\bibfnamefont {C.}~\bibnamefont {Rigetti}}, \bibinfo {author} {\bibfnamefont
  {S.}~\bibnamefont {Poletto}}, \bibinfo {author} {\bibfnamefont {G.~A.}\
  \bibnamefont {Keefe}}, \bibinfo {author} {\bibfnamefont {M.~B.}\ \bibnamefont
  {Rothwell}}, \bibinfo {author} {\bibfnamefont {J.~R.}\ \bibnamefont {Rozen}},
  \bibinfo {author} {\bibfnamefont {M.~B.}\ \bibnamefont {Ketchen}}, \ and\
  \bibinfo {author} {\bibfnamefont {M.}~\bibnamefont {Steffen}},\ }\bibfield
  {title} {\enquote {\bibinfo {title} {{Universal Quantum Gate Set Approaching
  Fault-Tolerant Thresholds with Superconducting Qubits}},}\ }\href {\doibase
  10.1103/PhysRevLett.109.060501} {\bibfield  {journal} {\bibinfo  {journal}
  {Phys. Rev. Lett.}\ }\textbf {\bibinfo {volume} {109}},\ \bibinfo {pages}
  {060501} (\bibinfo {year} {2012})}\BibitemShut {NoStop}%
\bibitem [{Note4()}]{Note4}%
  \BibitemOpen
  \bibinfo {note} {Eq.~\protect \textup {\hbox {\mathsurround \z@ \protect
  \normalfont (\ignorespaces \ref {eq:RWA2}\unskip \@@italiccorr )}} is in a
  rotated basis compared to \protect {$\sigma _z^{\protect \text {L}}\sigma
  _z^{\protect \text {R}}$}, but it generates the equivalent entangling gate.
  It is well known that \protect {$\sigma _z^{\protect \text {L}}\sigma
  _z^{\protect \text {R}}$} is maximally entangling with \begin {align*}
  e^{-i\protect \frac {\pi }{4}\sigma _z^{\protect \text {L}}\sigma
  _z^{\protect \text {R}}}= e^{i\protect \frac {\pi }{4}} e^{-i\protect \frac
  {5\pi }{4}\sigma _z^{\protect \text {L}}} e^{-i\protect \frac {5\pi
  }{4}\sigma _z^{\protect \text {R}}} \protect \text {CPHASE}. \end {align*}
  CPHASE has the Makhlin invariants\cite {makhlin2002} $G_1=0$ and
  $G_2=1$.}\BibitemShut {Stop}%
\bibitem [{\citenamefont {Burkard}\ \emph {et~al.}(1999)\citenamefont
  {Burkard}, \citenamefont {Loss},\ and\ \citenamefont
  {DiVincenzo}}]{burkard1999}%
  \BibitemOpen
  \bibfield  {author} {\bibinfo {author} {\bibfnamefont {G.}~\bibnamefont
  {Burkard}}, \bibinfo {author} {\bibfnamefont {D.}~\bibnamefont {Loss}}, \
  and\ \bibinfo {author} {\bibfnamefont {D.~P.}\ \bibnamefont {DiVincenzo}},\
  }\bibfield  {title} {\enquote {\bibinfo {title} {Coupled quantum dots as
  quantum gates},}\ }\href {\doibase 10.1103/PhysRevB.59.2070} {\bibfield
  {journal} {\bibinfo  {journal} {Phys. Rev. B}\ }\textbf {\bibinfo {volume}
  {59}},\ \bibinfo {pages} {2070} (\bibinfo {year} {1999})}\BibitemShut
  {NoStop}%
\bibitem [{\citenamefont {Hanson}\ \emph {et~al.}(2007)\citenamefont {Hanson},
  \citenamefont {Kouwenhoven}, \citenamefont {Petta}, \citenamefont {Tarucha},\
  and\ \citenamefont {Vandersypen}}]{hanson2007-2}%
  \BibitemOpen
  \bibfield  {author} {\bibinfo {author} {\bibfnamefont {R.}~\bibnamefont
  {Hanson}}, \bibinfo {author} {\bibfnamefont {L.~P.}\ \bibnamefont
  {Kouwenhoven}}, \bibinfo {author} {\bibfnamefont {J.~R.}\ \bibnamefont
  {Petta}}, \bibinfo {author} {\bibfnamefont {S.}~\bibnamefont {Tarucha}}, \
  and\ \bibinfo {author} {\bibfnamefont {L.~M.~K.}\ \bibnamefont
  {Vandersypen}},\ }\bibfield  {title} {\enquote {\bibinfo {title} {Spins in
  few-electron quantum dots},}\ }\href {\doibase 10.1103/RevModPhys.79.1217}
  {\bibfield  {journal} {\bibinfo  {journal} {Rev. Mod. Phys.}\ }\textbf
  {\bibinfo {volume} {79}},\ \bibinfo {pages} {1217} (\bibinfo {year}
  {2007})}\BibitemShut {NoStop}%
\bibitem [{\citenamefont {Taylor}\ \emph {et~al.}(2007)\citenamefont {Taylor},
  \citenamefont {Petta}, \citenamefont {Johnson}, \citenamefont {Yacoby},
  \citenamefont {Marcus},\ and\ \citenamefont {Lukin}}]{taylor2007}%
  \BibitemOpen
  \bibfield  {author} {\bibinfo {author} {\bibfnamefont {J.~M.}\ \bibnamefont
  {Taylor}}, \bibinfo {author} {\bibfnamefont {J.~R.}\ \bibnamefont {Petta}},
  \bibinfo {author} {\bibfnamefont {A.~C.}\ \bibnamefont {Johnson}}, \bibinfo
  {author} {\bibfnamefont {A.}~\bibnamefont {Yacoby}}, \bibinfo {author}
  {\bibfnamefont {C.~M.}\ \bibnamefont {Marcus}}, \ and\ \bibinfo {author}
  {\bibfnamefont {M.~D.}\ \bibnamefont {Lukin}},\ }\bibfield  {title} {\enquote
  {\bibinfo {title} {Relaxation, dephasing, and quantum control of electron
  spins in double quantum dots},}\ }\href {\doibase 10.1103/PhysRevB.76.035315}
  {\bibfield  {journal} {\bibinfo  {journal} {Phys. Rev. B}\ }\textbf {\bibinfo
  {volume} {76}},\ \bibinfo {pages} {035315} (\bibinfo {year}
  {2007})}\BibitemShut {NoStop}%
\bibitem [{\citenamefont {Coish}\ and\ \citenamefont
  {Baugh}(2009)}]{coish2009}%
  \BibitemOpen
  \bibfield  {author} {\bibinfo {author} {\bibfnamefont {W.~A.}\ \bibnamefont
  {Coish}}\ and\ \bibinfo {author} {\bibfnamefont {J.}~\bibnamefont {Baugh}},\
  }\bibfield  {title} {\enquote {\bibinfo {title} {Nuclear spins in
  nanostructures},}\ }\href {\doibase 10.1002/pssb.200945229} {\bibfield
  {journal} {\bibinfo  {journal} {Phys. Status Solidi B}\ }\textbf {\bibinfo
  {volume} {246}},\ \bibinfo {pages} {2203} (\bibinfo {year}
  {2009})}\BibitemShut {NoStop}%
\bibitem [{\citenamefont {Medford}\ \emph
  {et~al.}(2013{\natexlab{b}})\citenamefont {Medford}, \citenamefont {Beil},
  \citenamefont {Taylor}, \citenamefont {Bartlett}, \citenamefont {Doherty},
  \citenamefont {Rashba}, \citenamefont {DiVincenzo}, \citenamefont {Lu},
  \citenamefont {Gossard},\ and\ \citenamefont {Marcus}}]{medford2013-2}%
  \BibitemOpen
  \bibfield  {author} {\bibinfo {author} {\bibfnamefont {J.}~\bibnamefont
  {Medford}}, \bibinfo {author} {\bibfnamefont {J.}~\bibnamefont {Beil}},
  \bibinfo {author} {\bibfnamefont {J.~M.}\ \bibnamefont {Taylor}}, \bibinfo
  {author} {\bibfnamefont {S.~D.}\ \bibnamefont {Bartlett}}, \bibinfo {author}
  {\bibfnamefont {A.~C.}\ \bibnamefont {Doherty}}, \bibinfo {author}
  {\bibfnamefont {E.~I.}\ \bibnamefont {Rashba}}, \bibinfo {author}
  {\bibfnamefont {D.~P.}\ \bibnamefont {DiVincenzo}}, \bibinfo {author}
  {\bibfnamefont {H.}~\bibnamefont {Lu}}, \bibinfo {author} {\bibfnamefont
  {A.~C.}\ \bibnamefont {Gossard}}, \ and\ \bibinfo {author} {\bibfnamefont
  {C.~M.}\ \bibnamefont {Marcus}},\ }\bibfield  {title} {\enquote {\bibinfo
  {title} {Self-consistent measurement and state tomography of an exchange-only
  spin qubit},}\ }\href {\doibase 10.1038/nnano.2013.168} {\bibfield  {journal}
  {\bibinfo  {journal} {Nat. Nanotechnol.}\ }\textbf {\bibinfo {volume} {8}},\
  \bibinfo {pages} {654} (\bibinfo {year} {2013}{\natexlab{b}})}\BibitemShut
  {NoStop}%
\bibitem [{\citenamefont {Fowler}\ \emph {et~al.}(2012)\citenamefont {Fowler},
  \citenamefont {Mariantoni}, \citenamefont {Martinis},\ and\ \citenamefont
  {Cleland}}]{fowler2012}%
  \BibitemOpen
  \bibfield  {author} {\bibinfo {author} {\bibfnamefont {A.~G.}\ \bibnamefont
  {Fowler}}, \bibinfo {author} {\bibfnamefont {M.}~\bibnamefont {Mariantoni}},
  \bibinfo {author} {\bibfnamefont {J.~M.}\ \bibnamefont {Martinis}}, \ and\
  \bibinfo {author} {\bibfnamefont {A.~N.}\ \bibnamefont {Cleland}},\
  }\bibfield  {title} {\enquote {\bibinfo {title} {{Surface codes: Towards
  practical large-scale quantum computation}},}\ }\href {\doibase
  10.1103/PhysRevA.86.032324} {\bibfield  {journal} {\bibinfo  {journal} {Phys.
  Rev. A}\ }\textbf {\bibinfo {volume} {86}},\ \bibinfo {pages} {032324}
  (\bibinfo {year} {2012})}\BibitemShut {NoStop}%
\bibitem [{\citenamefont {Jones}\ \emph {et~al.}(2012)\citenamefont {Jones},
  \citenamefont {{Van Meter}}, \citenamefont {Fowler}, \citenamefont {McMahon},
  \citenamefont {Kim}, \citenamefont {Ladd},\ and\ \citenamefont
  {Yamamoto}}]{jones2012}%
  \BibitemOpen
  \bibfield  {author} {\bibinfo {author} {\bibfnamefont {N.~C.}\ \bibnamefont
  {Jones}}, \bibinfo {author} {\bibfnamefont {R.}~\bibnamefont {{Van Meter}}},
  \bibinfo {author} {\bibfnamefont {A.~G.}\ \bibnamefont {Fowler}}, \bibinfo
  {author} {\bibfnamefont {P.~L.}\ \bibnamefont {McMahon}}, \bibinfo {author}
  {\bibfnamefont {J.}~\bibnamefont {Kim}}, \bibinfo {author} {\bibfnamefont
  {T.~D.}\ \bibnamefont {Ladd}}, \ and\ \bibinfo {author} {\bibfnamefont
  {Y.}~\bibnamefont {Yamamoto}},\ }\bibfield  {title} {\enquote {\bibinfo
  {title} {{Layered Architecture for Quantum Computing}},}\ }\href {\doibase
  10.1103/PhysRevX.2.031007} {\bibfield  {journal} {\bibinfo  {journal} {Phys.
  Rev. X}\ }\textbf {\bibinfo {volume} {2}},\ \bibinfo {pages} {031007}
  (\bibinfo {year} {2012})}\BibitemShut {NoStop}%
\bibitem [{\citenamefont {Barthel}\ \emph
  {et~al.}(2010{\natexlab{b}})\citenamefont {Barthel}, \citenamefont {Medford},
  \citenamefont {Marcus}, \citenamefont {Hanson},\ and\ \citenamefont
  {Gossard}}]{barthel2010-1}%
  \BibitemOpen
  \bibfield  {author} {\bibinfo {author} {\bibfnamefont {C.}~\bibnamefont
  {Barthel}}, \bibinfo {author} {\bibfnamefont {J.}~\bibnamefont {Medford}},
  \bibinfo {author} {\bibfnamefont {C.~M.}\ \bibnamefont {Marcus}}, \bibinfo
  {author} {\bibfnamefont {M.~P.}\ \bibnamefont {Hanson}}, \ and\ \bibinfo
  {author} {\bibfnamefont {A.~C.}\ \bibnamefont {Gossard}},\ }\bibfield
  {title} {\enquote {\bibinfo {title} {{Interlaced Dynamical Decoupling and
  Coherent Operation of a Singlet-Triplet Qubit}},}\ }\href {\doibase
  10.1103/PhysRevLett.105.266808} {\bibfield  {journal} {\bibinfo  {journal}
  {Phys. Rev. Lett.}\ }\textbf {\bibinfo {volume} {105}},\ \bibinfo {pages}
  {266808} (\bibinfo {year} {2010}{\natexlab{b}})}\BibitemShut {NoStop}%
\bibitem [{\citenamefont {Bluhm}\ \emph {et~al.}(2011)\citenamefont {Bluhm},
  \citenamefont {Foletti}, \citenamefont {Neder}, \citenamefont {Rudner},
  \citenamefont {Mahalu}, \citenamefont {Umansky},\ and\ \citenamefont
  {Yacoby}}]{bluhm2011}%
  \BibitemOpen
  \bibfield  {author} {\bibinfo {author} {\bibfnamefont {H}~\bibnamefont
  {Bluhm}}, \bibinfo {author} {\bibfnamefont {S.}~\bibnamefont {Foletti}},
  \bibinfo {author} {\bibfnamefont {I.}~\bibnamefont {Neder}}, \bibinfo
  {author} {\bibfnamefont {M.}~\bibnamefont {Rudner}}, \bibinfo {author}
  {\bibfnamefont {D.}~\bibnamefont {Mahalu}}, \bibinfo {author} {\bibfnamefont
  {V.}~\bibnamefont {Umansky}}, \ and\ \bibinfo {author} {\bibfnamefont
  {A.}~\bibnamefont {Yacoby}},\ }\bibfield  {title} {\enquote {\bibinfo {title}
  {{Dephasing time of GaAs electron-spin qubits coupled to a nuclear bath
  exceeding $200~\mu\text{s}$}},}\ }\href {\doibase 10.1038/nphys1856}
  {\bibfield  {journal} {\bibinfo  {journal} {Nat. Phys.}\ }\textbf {\bibinfo
  {volume} {7}},\ \bibinfo {pages} {109} (\bibinfo {year} {2011})}\BibitemShut
  {NoStop}%
\bibitem [{\citenamefont {Medford}\ \emph {et~al.}(2012)\citenamefont
  {Medford}, \citenamefont {Cywinski}, \citenamefont {Barthel}, \citenamefont
  {Marcus}, \citenamefont {Hanson},\ and\ \citenamefont
  {Gossard}}]{medford2012}%
  \BibitemOpen
  \bibfield  {author} {\bibinfo {author} {\bibfnamefont {J.}~\bibnamefont
  {Medford}}, \bibinfo {author} {\bibfnamefont {L.}~\bibnamefont {Cywinski}},
  \bibinfo {author} {\bibfnamefont {C.}~\bibnamefont {Barthel}}, \bibinfo
  {author} {\bibfnamefont {C.~M.}\ \bibnamefont {Marcus}}, \bibinfo {author}
  {\bibfnamefont {M.~P.}\ \bibnamefont {Hanson}}, \ and\ \bibinfo {author}
  {\bibfnamefont {A.~C.}\ \bibnamefont {Gossard}},\ }\bibfield  {title}
  {\enquote {\bibinfo {title} {{Scaling of Dynamical Decoupling for Spin
  Qubits}},}\ }\href {\doibase 10.1103/PhysRevLett.108.086802} {\bibfield
  {journal} {\bibinfo  {journal} {Phys. Rev. Lett.}\ }\textbf {\bibinfo
  {volume} {108}},\ \bibinfo {pages} {086802} (\bibinfo {year}
  {2012})}\BibitemShut {NoStop}%
\bibitem [{\citenamefont {Foletti}\ \emph {et~al.}(2009)\citenamefont
  {Foletti}, \citenamefont {Bluhm}, \citenamefont {Mahalu}, \citenamefont
  {Umansky},\ and\ \citenamefont {Yacoby}}]{foletti2009}%
  \BibitemOpen
  \bibfield  {author} {\bibinfo {author} {\bibfnamefont {S.}~\bibnamefont
  {Foletti}}, \bibinfo {author} {\bibfnamefont {H.}~\bibnamefont {Bluhm}},
  \bibinfo {author} {\bibfnamefont {D.}~\bibnamefont {Mahalu}}, \bibinfo
  {author} {\bibfnamefont {V.}~\bibnamefont {Umansky}}, \ and\ \bibinfo
  {author} {\bibfnamefont {A.}~\bibnamefont {Yacoby}},\ }\bibfield  {title}
  {\enquote {\bibinfo {title} {Universal quantum control of two-electron spin
  quantum bits using dynamic nuclear polarization},}\ }\href {\doibase
  10.1038/nphys1424} {\bibfield  {journal} {\bibinfo  {journal} {Nat. Phys.}\
  }\textbf {\bibinfo {volume} {5}},\ \bibinfo {pages} {903} (\bibinfo {year}
  {2009})}\BibitemShut {NoStop}%
\bibitem [{\citenamefont {Bluhm}\ \emph {et~al.}(2010)\citenamefont {Bluhm},
  \citenamefont {Foletti}, \citenamefont {Mahalu}, \citenamefont {Umansky},\
  and\ \citenamefont {Yacoby}}]{bluhm2010}%
  \BibitemOpen
  \bibfield  {author} {\bibinfo {author} {\bibfnamefont {H.}~\bibnamefont
  {Bluhm}}, \bibinfo {author} {\bibfnamefont {S.}~\bibnamefont {Foletti}},
  \bibinfo {author} {\bibfnamefont {D.}~\bibnamefont {Mahalu}}, \bibinfo
  {author} {\bibfnamefont {V.}~\bibnamefont {Umansky}}, \ and\ \bibinfo
  {author} {\bibfnamefont {A.}~\bibnamefont {Yacoby}},\ }\bibfield  {title}
  {\enquote {\bibinfo {title} {{Enhancing the Coherence of a Spin Qubit by
  Operating it as a Feedback Loop That Controls its Nuclear Spin Bath}},}\
  }\href {\doibase 10.1103/PhysRevLett.105.216803} {\bibfield  {journal}
  {\bibinfo  {journal} {Phys. Rev. Lett.}\ }\textbf {\bibinfo {volume} {105}},\
  \bibinfo {pages} {216803} (\bibinfo {year} {2010})}\BibitemShut {NoStop}%
\bibitem [{\citenamefont {Veldhorst}\ \emph
  {et~al.}(2014{\natexlab{b}})\citenamefont {Veldhorst}, \citenamefont {Yang},
  \citenamefont {Hwang}, \citenamefont {Huang}, \citenamefont {Dehollain},
  \citenamefont {Muhonen}, \citenamefont {Simmons}, \citenamefont {Laucht},
  \citenamefont {Hudson}, \citenamefont {Itoh}, \citenamefont {Morello},\ and\
  \citenamefont {Dzurak}}]{veldhorst2014-2}%
  \BibitemOpen
  \bibfield  {author} {\bibinfo {author} {\bibfnamefont {M.}~\bibnamefont
  {Veldhorst}}, \bibinfo {author} {\bibfnamefont {C.~H.}\ \bibnamefont {Yang}},
  \bibinfo {author} {\bibfnamefont {J.~C.~C.}\ \bibnamefont {Hwang}}, \bibinfo
  {author} {\bibfnamefont {W.}~\bibnamefont {Huang}}, \bibinfo {author}
  {\bibfnamefont {J.~P.}\ \bibnamefont {Dehollain}}, \bibinfo {author}
  {\bibfnamefont {J.~T.}\ \bibnamefont {Muhonen}}, \bibinfo {author}
  {\bibfnamefont {S.}~\bibnamefont {Simmons}}, \bibinfo {author} {\bibfnamefont
  {A.}~\bibnamefont {Laucht}}, \bibinfo {author} {\bibfnamefont {F.~E.}\
  \bibnamefont {Hudson}}, \bibinfo {author} {\bibfnamefont {K.~M.}\
  \bibnamefont {Itoh}}, \bibinfo {author} {\bibfnamefont {A.}~\bibnamefont
  {Morello}}, \ and\ \bibinfo {author} {\bibfnamefont {A.~S.}\ \bibnamefont
  {Dzurak}},\ }\bibfield  {title} {\enquote {\bibinfo {title} {{A Two Qubit
  Logic Gate in Silicon}},}\ }\href {http://lanl.arxiv.org/abs/1411.5760}
  {\bibfield  {journal} {\bibinfo  {journal} {arXiv:1411.5760
  [cond-mat.mes-hall]}\ } (\bibinfo {year} {2014}{\natexlab{b}})}\BibitemShut
  {NoStop}%
\bibitem [{\citenamefont {Kim}\ \emph {et~al.}(2015{\natexlab{b}})\citenamefont
  {Kim}, \citenamefont {Ward}, \citenamefont {Simmons}, \citenamefont {Savage},
  \citenamefont {Lagally}, \citenamefont {Friesen}, \citenamefont
  {Coppersmith},\ and\ \citenamefont {Eriksson}}]{kim2015}%
  \BibitemOpen
  \bibfield  {author} {\bibinfo {author} {\bibfnamefont {D.}~\bibnamefont
  {Kim}}, \bibinfo {author} {\bibfnamefont {D.~R.}\ \bibnamefont {Ward}},
  \bibinfo {author} {\bibfnamefont {C.~B.}\ \bibnamefont {Simmons}}, \bibinfo
  {author} {\bibfnamefont {D.~E.}\ \bibnamefont {Savage}}, \bibinfo {author}
  {\bibfnamefont {M.~G.}\ \bibnamefont {Lagally}}, \bibinfo {author}
  {\bibfnamefont {M.}~\bibnamefont {Friesen}}, \bibinfo {author} {\bibfnamefont
  {S.~N.}\ \bibnamefont {Coppersmith}}, \ and\ \bibinfo {author} {\bibfnamefont
  {M.~A.}\ \bibnamefont {Eriksson}},\ }\bibfield  {title} {\enquote {\bibinfo
  {title} {High fidelity resonant gating of a silicon based quantum dot hybrid
  qubit},}\ }\href {http://lanl.arxiv.org/abs/1502.03156} {\bibfield  {journal}
  {\bibinfo  {journal} {arXiv:1502.03156 [cond-mat.mes-hall]}\ } (\bibinfo
  {year} {2015}{\natexlab{b}})}\BibitemShut {NoStop}%
\bibitem [{\citenamefont {Nielsen}\ and\ \citenamefont
  {Chuang}(2000)}]{nielsen2000}%
  \BibitemOpen
  \bibfield  {author} {\bibinfo {author} {\bibfnamefont {M.~A.}\ \bibnamefont
  {Nielsen}}\ and\ \bibinfo {author} {\bibfnamefont {I.~L.}\ \bibnamefont
  {Chuang}},\ }\href@noop {} {\emph {\bibinfo {title} {Quantum Computation and
  Quantum Information}}}\ (\bibinfo  {publisher} {Cambridge University Press},\
  \bibinfo {address} {Cambridge},\ \bibinfo {year} {2000})\BibitemShut
  {NoStop}%
\end{thebibliography}%
\end{document}